\documentclass[isre,nonblindrev]{informs3} 
\DoubleSpacedXI

\usepackage{natbib}
 \bibpunct[, ]{(}{)}{,}{a}{}{,}
 \def\bibfont{\small}

\usepackage{float}
\usepackage{makecell}
\usepackage[para]{threeparttable}
\usepackage{verbatim}

\usepackage{algorithm}
\usepackage{algpseudocode}

\usepackage{booktabs}

\usepackage{enumitem}

\setlist[enumerate]{
  itemsep=6pt,      
  topsep=6pt,       
  leftmargin=1.2cm, 
  labelsep=0.4em,   
  align=left
}

\newcommand{\surveysection}[1]{\vspace{0.8em}\textit{#1}\par}

\usepackage[justification=centering,singlelinecheck=false]{caption}

\usepackage[T1]{fontenc}

\usepackage[whole]{bxcjkjatype} 

\TheoremsNumberedThrough    
\ECRepeatTheorems

\EquationsNumberedThrough
\MANUSCRIPTNO{ISRE-0000-0000.00}

\renewcommand{\theARTICLETOP}{}
\begin{document}

\RUNTITLE{Generative Search Engine}
\TITLE{When Content is Goliath and Algorithm is David: The Style and Semantic Effects of Generative Search Engine}

\ARTICLEAUTHORS{
\AUTHOR{Lijia Ma\thanks{These authors contributed equally to the manuscript and are listed alphabetically.}}
\AFF{Belk College of Business, University of North Carolina at Charlotte}
\AUTHOR{Juan Qin\footnotemark[2]}
\AFF{Faculty of Business for Science and Technology, School of Management, University of Science and Technology of China} 
\AUTHOR{Xingchen (Cedric) Xu\footnotemark[2], Yong Tan\thanks{Corresponding author.}}
\AFF{Michael G. Foster School of Business, University of Washington} 
} 

\ABSTRACT{Generative search engines (GEs) leverage large language models (LLMs) to deliver AI-generated summaries with website citations, establishing novel traffic acquisition channels while fundamentally altering the search engine optimization landscape. To investigate the distinctive characteristics of GEs, we collect data through interactions with Google's generative and conventional search platforms, compiling a dataset of approximately ten thousand websites across both channels. Our empirical analysis reveals that GEs exhibit preferences for citing content characterized by significantly higher predictability for underlying LLMs and greater semantic similarity among selected sources. Through controlled experiments utilizing retrieval augmented generation (RAG) APIs, we demonstrate that these citation preferences emerge from intrinsic LLM tendencies to favor content aligned with their generative expression patterns. Motivated by applications of LLMs to optimize website content, we conduct additional experimentation to explore how LLM-based content polishing by website proprietors alters AI summaries, finding that such polishing paradoxically enhances information diversity within AI summaries. Finally, to assess the user-end impact of LLM-induced information increases, we design a generative search engine and recruit Prolific participants to conduct a randomized controlled experiment involving an information-seeking and writing task. We find that higher-educated users exhibit minimal changes in their final outputs' information diversity but demonstrate significantly reduced task completion time when original sites undergo polishing. Conversely, lower-educated users primarily benefit through enhanced information density in their task outputs while maintaining similar completion times across experimental groups. Our findings provide theoretical and practical implications for stakeholders in the evolving search ecosystem.
}


\maketitle
\newpage

``The medium is the message.'' - \cite{mcluhan1964understanding}

\section{Introduction}

In the contemporary digital ecosystem, the exponential growth of information volume and diversity has necessitated sophisticated intermediary mechanisms to facilitate connections between information providers and consumers. Search engines have emerged as critical infrastructure in this information mediation process, becoming integral to daily digital interactions. Empirical evidence indicates that 87\% of mobile device users utilize search engines at least once daily, while Google alone processes approximately 63,000 queries per second, corresponding to roughly 2 trillion annual searches globally\footnote{See \url{https://serpwatch.io/blog/search-engine-statistics/}.}. These platforms algorithmically determine content visibility through complex scoring mechanisms that evaluate query-content relevance and numerous additional ranking factors. The complexity and opacity of these algorithmic systems have catalyzed the emergence of Search Engine Optimization (SEO) as a specialized industry, which encompasses systematic strategies for deciphering algorithmic ranking criteria and optimizing digital content to enhance organic search visibility \citep{liu2018semantic}. This sector has demonstrated substantial economic significance, achieving a market valuation of US\$89.1 billion in 2024\footnote{See \url{https://www.researchandmarkets.com/reports/5140303/search-engine-optimization-seo-global}.}.

However, the search engine paradigm is undergoing a fundamental transformation driven by the emergence of \textit{generative search engines} (GE) that leverage Large Language Models (LLMs) and Retrieval Augmented Generation (RAG) architectures. This technological shift was catalyzed by OpenAI's introduction of ChatGPT in November 2022, which pioneered conversational query resolution mechanisms. Microsoft subsequently integrated GPT-4 technology into its search infrastructure, launching the New Bing platform in February 2023\footnote{See \url{https://techcrunch.com/2023/03/14/microsofts-new-bing-was-using-gpt-4-all-along/}.}. Later, Google also initiated experimental deployment of generative search capabilities in March 2023, which evolved into the ``\textit{AI Overviews}'' feature, an AI-powered summarization system integrated into search results, launched in May 2024\footnote{See \url{https://blog.google/products/search/generative-ai-google-search-may-2024/}}. The demonstrated efficacy of these systems in enhancing search efficiency and user satisfaction has facilitated rapid adoption and geographic expansion, with AI Overviews subsequently deployed across more than 200 countries and engaging 2 billion monthly users as of July 2025\footnote{See \url{https://blog.google/inside-google/message-ceo/alphabet-earnings-q2-2025/}}. This paradigmatic evolution represents a fundamental departure from conventional search engines, which primarily function as indexing systems that enumerate relevant websites; instead, generative search engines employ sophisticated query interpretation mechanisms, dynamically synthesize information from multiple sources, and generate coherent responses that incorporate citations for verification and attribution.

Extant research on conventional search engines has demonstrated the substantial influence of search engine visibility on individual decision-making processes \citep{gong2018examining, ghose2014examining} and the distribution of welfare among market stakeholders \citep{berman2013role}. As the search paradigm increasingly incorporates generative search engines, website visibility has become contingent upon citation frequency within LLM-generated content summaries and overviews. This fundamental shift in visibility determinants necessitates strategic adaptation beyond traditional SEO methodologies; websites pursuing enhanced exposure must now develop specialized approaches to optimize their citation probability within generative engine outputs. Consequently, this technological evolution has led to the emergence of a nascent industry termed Generative Engine Optimization (GEO), which focuses specifically on maximizing website citation rates in AI-generated search responses. However, optimization across both conventional and generative search paradigms presents significant challenges due to the opaque nature of underlying algorithms to the public, deliberately designed to maintain system integrity and reduce susceptibility to manipulation \citep{erdmann2022search, danaher2006factors}. Scholarly investigations in this domain have employed Natural Language Processing (NLP) methodologies to examine the relationship between various content factors and website rankings in search engine results \citep{liu2018semantic}, subsequently informing content optimization strategies \citep{reisenbichler2022frontiers}. However, GEO presents even greater difficulties due to the vast output space and randomness of their underlying large language models, whose behavioral patterns remain only partially predictable even to the algorithm developers and system architects who designed them \citep{berti2025emergent}. Therefore, we ask our first research question:

\textit{RQ1: How do generative search engines select websites to cite in response to a query?} 

To explore this research question, we construct a large-scale dataset encompassing over ten thousand websites through interaction with Google's search infrastructure, capturing AI Overview content, cited sources within AI Overviews, and top-ranked websites in conventional Google Search results. Our analysis reveals that AI Overview predominantly cites content exhibiting significantly lower perplexity compared to content prioritized by conventional search algorithms. This phenomenon stems from the inherent characteristics of the Transformer architecture employed by most contemporary LLMs, including Google's underlying Gemini model, which operates through auto-regressive token generation mechanisms \citep{vaswani2017attention}. Specifically, these models generate natural language responses sequentially, selecting tokens with high conditional probability at each generation step. Consequently, when LLMs synthesize AI Overview content (hereafter referred to as ``AI summary'') and incorporate textual elements from source materials, the linguistic predictability of these original sources, as evaluated by the LLM's internal probability distributions, directly influences the overall predictability of the generated AI summary, which the model seeks to optimize during the generation process. Notably, perplexity measurements demonstrate minimal influence on ranking within conventional Google search results, highlighting a fundamental divergence in algorithmic prioritization mechanisms. Furthermore, our analysis demonstrates that content cited by LLMs exhibits greater semantic homogeneity compared to sources ranked highly by conventional search engines. This semantic convergence occurs because AI Overviews are architecturally designed to synthesize coherent, unified responses rather than merely aggregating disparate information sources. Consequently, despite the potential for generative engines to incorporate diverse, niche websites that do not achieve prominent rankings in conventional search results, the cited sources consistently demonstrate substantial semantic similarity due to the coherence requirements inherent in the generative summarization process.

Nevertheless, it remains uncertain whether these observed citation criteria represent explicit engineering decisions specifically tailored for Google's AI Overview functionality or constitute inherent characteristics embedded within the underlying language model architecture. Two competing explanations emerge from this uncertainty. First, Google's AI Overview confronts the substantial challenge of processing vast quantities of dynamic web content, potentially necessitating additional algorithmic layers and engineered preferences that might lead to specific content characteristics. Alternatively, following our preceding theoretical framework, the observed stylistic preferences (reflected in perplexity patterns) may derive from the fundamental operational mechanisms of LLMs (specifically, their next-token prediction paradigm), while the semantic tendencies (manifested as content similarity) may emerge from the inherent constraints of the output format requirement for coherent natural language summarization. Given these competing perspectives, we pose our second research question:

\textit{RQ2: Do generative search engines' criteria originate through manual curation or naturally emerge from underlying language models?}

To address this research inquiry, we investigate the information retrieval and content synthesis processes executed by Gemini, the foundational model architecture underlying Google's AI Overview functionality. If the observed preferences for citation source selection are attributable to intrinsic model characteristics rather than explicit engineering interventions, analogous patterns should manifest in independent implementations utilizing Gemini for Retrieval Augmented Generation (RAG) tasks. To test this hypothesis, we construct an additional dataset using Gemini's document understanding API by providing both search queries and curated compilations of website content as inputs. Replicating our analyses within this controlled environment, we discover that the RAG system exhibits qualitatively consistent stylistic and semantic properties, thereby corroborating the generalizability of our findings and supporting the hypothesis that these citation preferences are intrinsic to the underlying language model architecture. Throughout this process, we also identify and empirically control for positional bias inherent in RAG systems, thereby generating additional insights for GEO practice.

Building on our empirical findings, we extend our investigation to examine another critical phenomenon in the digital content landscape. The proliferation of LLMs has introduced widespread adoption of automated content creation and refinement processes, with the AI-generated content market reaching a valuation of \$13.9 billion as of 2024\footnote{See \url{https://www.grandviewresearch.com/industry-analysis/generative-ai-content-creation-market-report}}. This technological shift has attracted considerable scholarly attention, as demonstrated by \cite{ye2025lola}'s LLM-based headline generation for traffic enhancement and \cite{reisenbichler2022frontiers}'s pioneering GPT-2 application for SEO optimization even before ChatGPT's introduction. In the age of GEO, our preceding analysis demonstrates that generative engines exhibit systematic preferences for linguistically predictable content, suggesting a natural optimization strategy: leveraging LLMs to refine existing content, thereby enhancing its predictability from the model's perspective. However, this approach raises a fundamental question regarding content diversity. Universal adoption of LLM-based content polishing may precipitate content similarity, a phenomenon recently documented in the literature as LLMs' homogenizing effect \citep{moon2024homogenizing}. Conversely, such AI-mediated refinement could potentially expand the consideration set for generative engines by transforming previously unpredictable content into more accessible forms, thereby enabling AI summaries to incorporate greater content diversity. To investigate these competing theoretical predictions, we pose our third research question:

\textit{RQ3: How does content diversity change when input content for generative search engines is refined through LLM-based polishing processes?}

To conduct this analysis, we implement a controlled RAG experiment following the identical methodological framework established for RQ2. However, prior to inputting content into Gemini's document understanding API, we deploy Gemini's text generation API to refine all source materials through automated polishing procedures. Our findings reveal an unexpected outcome: automated content refinement actually enhances information diversity of the RAG outputs, indicating that the diversity-expansion mechanism predominates over the homogenization effect. This mechanism receives further empirical substantiation as we observe an expanded citation scope encompassing a greater number of source websites. Additionally, our analysis demonstrates that explicitly incorporating citation optimization objectives within the AI polishing prompts significantly amplifies these diversity-enhancing effects, thereby revealing an intrinsic AI-to-AI compatibility phenomenon wherein language models inherently understand and accommodate the preferences of similar architectures, even in the absence of detailed optimization guidance.

Although our experimental results demonstrate that LLM-based content polishing significantly alters both the information contained in generated summaries and the websites cited therein, the effects of these content modifications on user behavior remain theoretically ambiguous. Two competing mechanisms emerge from this uncertainty. First, leveraging LLMs' natural language generation capabilities, AI Overviews deliver coherent paragraph-form responses that can enhance users' search efficiency \citep{xu2023chatgpt}, suggesting that increased information diversity may also be readily processed and utilized by users. Conversely, users remain constrained by limited attention and cognitive processing capacity, potentially limiting their ability to consume expanded information sets, a phenomenon documented as information overload \citep{oreilly1980individuals}. Consequently, despite enhanced information diversity resulting from content optimization, users may be unable to effectively process and leverage this additional information. Furthermore, as an emergent technological artifact, generative search systems may exhibit differential usage patterns and outcomes across different users, reflecting what scholars term the ``AI divide'' phenomenon \citep{ma2024learning, mcelheran2024ai}. Given these theoretical tensions, we pose our fourth research question:

\textit{RQ4: How do users' search experiences change when input content undergoes LLM-based polishing processes? Do such changes vary systematically across different users?}

To investigate these directions, we develop a platform that replicates Google AI Overview's functionality and recruit participants from Prolific to complete writing tasks using our system. We employ a randomized controlled trial design wherein half of the participants (control group) receive AI summaries generated from unmodified website content sourced directly from the internet, while the remaining participants (treatment group) receive summaries based on LLM-polished website content. Our experimental results yield multifaceted insights. First, corroborating our findings from RQ3, we observe significant increases in both information entropy and the number of cited websites within AI responses provided to treatment group participants. Second, this enhanced information diversity in AI Overview outputs translates directly to participants' performance outcomes: treatment group participants produce longer and more substantively informative task solutions. Third, and most unexpectedly, we identify differential treatment effects across educational backgrounds. Participants with undergraduate education or below derive primary benefits through improved solution quality, while those with graduate-level education experience efficiency gains rather than quality improvements. This differential response pattern emerges because graduate-educated participants exhibit adaptive prompting behavior, issuing additional queries when information density appears insufficient. When exposed to optimized websites containing enhanced information density, these participants reduce their query frequency, thereby achieving time savings while maintaining solutions' information density. Conversely, participants with undergraduate education or below demonstrate consistent low-prompting behavior across both experimental conditions, but directly benefit from the increased information content available in the treatment group AI summaries.

The remainder of this paper is organized as follows. Section 2 synthesizes relevant literature and situates our research within the existing scholarly framework. Section 3 delineates our approach for observational data collection and processing. Section 4 presents our variable construction procedures and empirical analyses, systematically addressing research questions 1-3. Section 5 describes our controlled experimental design involving human participants and corresponding analyses to examine research question 4. Section 6 evaluates the robustness of our findings through comprehensive validation procedures, including a replication study conducted on Microsoft's New Bing platform. Finally, Section 7 concludes by discussing our theoretical and practical contributions, examining broader implications for the search engine ecosystem, and proposing directions for future scholarly investigation.

\section{Related Literature}
Our research builds upon and contributes to four streams of literature: (i) search engine optimization and marketing; (ii) economics of information retrieval systems; (iii) AI-generated summaries and human-AI interaction.

\subsection{Search Engine Optimization and Marketing}

Contextually, our study directly relates to the search engine optimization and marketing literature. Given that search engine appearances and rankings profoundly influence consumer behavior and market outcomes \citep{ghose2014examining, goldfarb2011online}, extensive efforts have been undertaken to enhance websites' visibility across both organic and sponsored search channels.

Regarding organic search, ranking algorithms are typically maintained as opaque by search engine platforms to minimize direct manipulation by website owners. Consequently, the search engine optimization (SEO) literature seeks to identify factors that influence organic traffic, thereby enabling content optimization to increase organic visibility. This research spans technical elements and site architecture \citep{danaher2006factors} to content semantics \citep{liu2018semantic}. Recently, researchers have explored the application of generative AI to optimize content style and semantics for enhanced exposure in organic listings \citep{reisenbichler2022frontiers}. Regarding sponsored search, search engine platforms provide paid traffic opportunities as an additional channel for website owners. Search engine marketing (SEM) research develops strategies for bidding on such paid traffic from multiple perspectives, including optimal bidding across keywords and positions \citep{abhishek2013optimal}, budget allocation under constraints \citep{shin2015keyword}, and keyword matching strategies \citep{du2017bidding}.

However, as generative search engines emerge as a novel channel for website exposure, understanding of their distinctive traffic allocation patterns remains limited, particularly regarding how citation patterns relate to foundation models. We aim to extend SEO literature to generative engine optimization (GEO) by investigating the factors governing generative engines' citation behavior and the underlying mechanisms.

\subsection{Economics of Information Retrieval Systems}

Our study also relates to the economics of information retrieval systems, which examines the impacts and strategic interactions among participants within various information retrieval platforms, including search engines and recommender systems \citep{zhou2024longitudinal, berman2013role}.

Information retrieval technologies emerge due to consumers' limited capacity to access and process comprehensive information sets, particularly when dealing with large-scale information environments \citep{ursu2025sequential}. Search engines curate a targeted subset of relevant websites for users, eliminating the need for indiscriminate browsing across countless web resources \citep{gong2018examining}. Since search engines directly influence user exposure to online content and consequently affect website revenue generation, these platforms substantially redistribute welfare among advertisers, consumers, and themselves \citep{berman2013role}. 

Previous research has documented strategic platform behavior. For instance, \cite{hagiu2011intermediaries} explain why intermediaries may redirect user searches for profit maximization, even at the expense of consumer welfare, while \cite{decorniere2014integration} analyze vertical integration in search markets, demonstrating that integrated platforms possess incentives to bias rankings toward their proprietary services. Regarding website owners, prior research demonstrates their strategic responses to search engine mechanisms aimed at enhancing visibility within budget constraints. \cite{berman2013role} show that when SEO investments achieve sufficient effectiveness, advertisers can improve organic visibility while reducing dependence on sponsored links. Similarly, \cite{baye2016seo} analyze how retailers' investments in site quality and brand awareness increase organic traffic both directly through consumer preferences for superior and well-established sites, and indirectly through search engines' favorable positioning of such sites.

As generative search engines emerge as novel intermediaries for information transmission, their potential economic implications represent a compelling and significant research direction. We aim to establish foundational insights for future economic modeling by analyzing generative engines' behavioral patterns and examining potential output variations when websites strategically optimize content.

\subsection{AI-Generated Summaries and Human-AI Interaction}

Generative search engines provide AI-generated summaries in response to search requests. Previous research on AI-generated summaries typically examines user responses to such summaries across various online platforms. For instance, \cite{koo2025ai} find that AI-generated review summaries on e-commerce platforms guide subsequent ratings and comments toward more positive evaluations, thereby reinforcing the advantages of already popular products. \cite{alavi2023ai} document that AI product review summaries reduce engagement with individual reviews and decrease thematic diversity, suggesting an anchoring effect toward uniform content. Conversely, \cite{yuan2024ai} find that AI-generated summaries serve as differentiating references on hotel review platforms, enhancing diversity and helpfulness of subsequent user content while ultimately boosting sales. Extending this inquiry to online video platforms \citep{ji2025designing}, \cite{kim2024less} find that AI video summaries significantly increase user engagement, particularly through increased comments and unique commenters for longer videos.

In the generative search engine context, users can freely prompt engines to generate varied AI summaries, making the human-AI interaction literature equally relevant. Previous research employs laboratory experiments to examine how users interact with generative AI to complete information-seeking tasks \citep{xu2023chatgpt}. Further, user interaction patterns and performance may vary significantly when engaging with generative AI \citep{gao2025pandora, yeverechyahu2024impact, qiao2023ai, wang2023human}. For example, \cite{ma2024learning} find that lower-educated users derive greater utility from generative AI but learn about this utility more slowly. Such disparities in human-AI interaction constitute an important research area known as the AI divide. For instance, \cite{yan2024creative} demonstrate that generative AI alters the balance between ideation and execution, amplifying inequality by favoring certain creators over others. Consistent with this finding, \cite{hou2024double} find that generative AI enhances divergent thinking for less experienced creators while reducing efficiency for experts without improving design quality.

As a significant AI-driven channel through which users obtain information, the informational value provided by generative search engines for different user groups remains understudied, particularly when information diversity is altered by supply-side optimization. We aim to address this gap and extend understanding of human-AI interaction within such systems.

\section{Raw Data Sources and Collection}
\label{sec:data}
To investigate generative engines' website citation patterns and compare with traditional search engines, we collect data through direct interaction with both search paradigms. This section delineates our approach to query specification, engine interaction protocols, and the collection of source materials highlighted by each system. The resulting dataset serves as the foundation for subsequent analyses addressing research questions 1, 2, and 3 in the following section.

\subsection{Query Selection and Sampling}

To conduct a comprehensive and representative analysis of generative search engine behavior, we require a diverse set of input queries that are amenable to natural language answering and thus likely to elicit generative responses. We select the Human ChatGPT Comparison Corpus (HC3) \citep{guo2023close} as our query source. The HC3 dataset, originally developed to evaluate AI-generated versus human-generated text, encompasses diverse user queries spanning multiple domains and complexity levels, rendering it an ideal foundation for examining generative search engine responses across varied information needs. Given the computational expense of processing each query through both search engines and retrieving associated websites, we randomly sample 5,000 unique queries from HC3. This query sample size naturally involves tens of thousands of websites (discussed in the subsequent subsection), providing sufficient scale for large-scale analyses.

\subsection{Responses from AI Overview and Conventional Search Engine}

\begin{figure}[!h]
\FIGURE{\includegraphics[width=4.25in]{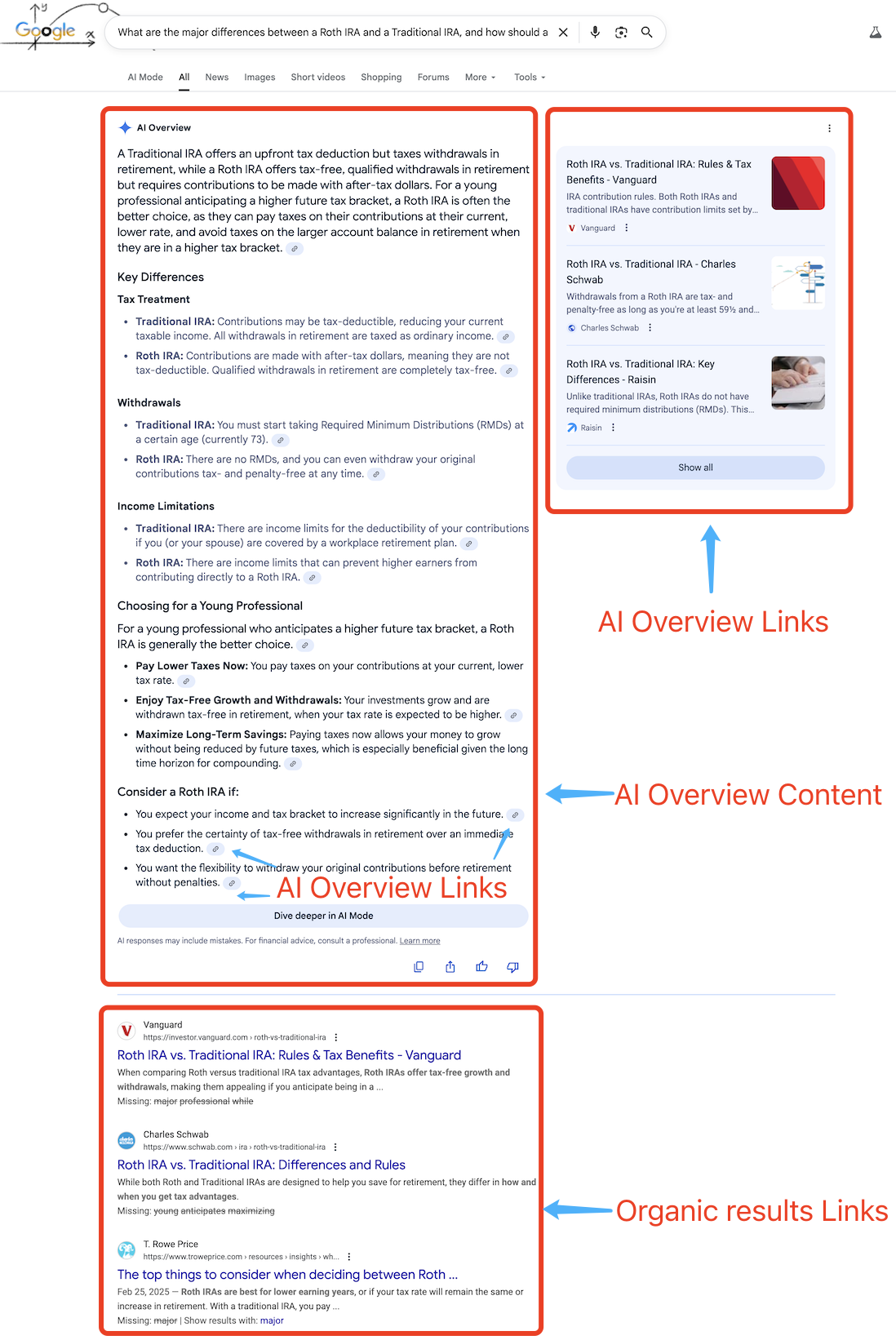}}
{Example of Google Search Results Page with AI Overview and Organic Results \label{img:google}}
{}
\end{figure}

Given Google's dominance in the search engine market and the widespread rollout of its AI Overview function to most countries in 2025, we select Google as our primary interaction target. As illustrated in Figure \ref{img:google}, Google's search interface presents organic website listings in response to user queries. When available, the platform also displays AI Overview results featuring in-text citations. The right-hand panel lists referenced websites, some of which correspond to these in-text citations.

We leverage SerpAPI, which provides reliable scraping services for Google search results including AI Overview functionality\footnote{See \url{https://serpapi.com}}, to obtain Google's responses in both conventional and generative forms during June 2025. SerpAPI offers two relevant endpoints: a general Google search endpoint that encompasses both conventional search results and AI Overview, and a dedicated Google AI Overview endpoint specifically for generative responses. We utilize the general endpoint to input each sampled query and retrieve: (i) the AI Overview content; (ii) URL links cited by the AI Overview; and (iii) the first page of conventional search results indicated by SerpAPI, including their ranking positions, titles, snippets, and URLs.

In certain cases, the Google AI Overview is not returned directly by the general endpoint; instead, a parameter in the endpoint output indicates that an additional request is required\footnote{See \url{https://serpapi.com/blog/scrape-google-ai-overviews/}}. For such queries, we subsequently invoke the AI Overview specific endpoint to retrieve the AI Overview content and its referenced URLs. This two-step procedure yields 4,060 queries containing both conventional and generative search results, which form the basis for our subsequent analyses. Among these, 305 queries provide AI summaries with references listed only at the end, whereas 3,755 include sentence-level references.

Following the retrieval of URLs from SerpAPI, we subsequently collect the complete content from all websites appearing in either AI Overview citations or conventional search results. We develop an automated web agent to dynamically extract textual information from these websites. Where websites appear in both search result types, we label them and de-duplicate the website content files, ultimately yielding 98,477 unique websites.

More details about the whole data collection process can be found in Appendix \ref{appendix:API}.

\section{Data Analyses}
\label{sec:main}

This section systematically examines research questions 1, 2, 3 and 4 through empirical analyses. Given that each research question necessitates distinct datasets and methodological frameworks, we adopt a uniform organizational structure for each subsection: the first part delineates the specific experimental design and variable construction procedures tailored to the respective research question, while the second part presents the corresponding empirical findings.

\subsection{Determinants of Website Citation in Generative Search Engines (RQ1)}
\label{sec: GEO_criteria}
\subsubsection{Chunk Selection and Variables}\

To investigate the criteria employed by generative search engines in selecting website content for citation, we construct a comprehensive dataset through systematic interaction with Google Search, as detailed in Section 3. For each query, we collect both AI Overview responses and the complete textual content of all websites appearing either in AI Overview citations or among the top conventional search results.

As previously noted, AI Overview comprises multiple sentences, each of which may cite zero to multiple websites as references, while individual websites in the AI Overview's reference list may be specifically cited by multiple sentences or alternatively listed without specific sentence attribution to provide general support for the AI summary's core content. Consequently, we can conduct empirical analyses at two distinct levels. First, we can evaluate whether a particular website merits citation for supporting the AI summary overall. Second, we can examine whether a specific website should be cited to support an individual sentence within the AI summary. However, given that some websites contain extensive information, most of which is not considered by the search engine when answering the user queries requested, we narrow the content scope accordingly.

For the first type of analysis, we consider three categories of websites and extract one representative chunk from each: websites cited by specific sentences in AI Overview, websites mentioned in AI Overview but not cited by specific sentences, and websites appearing only in conventional search results. For each category, we employ different matching targets as specified in Algorithm \ref{alg:representative_chunk}, while consistently using a rolling window approach to identify the chunk most relevant to the respective matching target. Specifically, we implement a sliding window approach with a window size of 128 characters and a step size of 16 characters to segment each website into overlapping chunks. We utilize the Sentence-BERT model \citep{reimers2019sentence} to generate embeddings for both matching targets and website chunks, subsequently computing cosine similarity to identify the most semantically representative chunk for each website.

\begin{algorithm}[h]
\caption{Chunk Selection: One Chunk per Website}
\label{alg:representative_chunk}
\begin{algorithmic}[1]
\Require AI Overview content, Search snippets, Website content
\For{each query $q$}
    \For{each website $k$ in query $q$}
        \If{website $k$ is cited by specific sentences in AI Overview}
            \State $S_k \gets$ all sentences citing website $k$
            \State $C_k \gets$ all chunks of website $k$ 
            \State Select $c^*_k = \arg\max_{c \in C_k; s \in S_k}$ similarity$(c, s)$
        \ElsIf{website $k$ is mentioned in AI Overview but not cited by specific sentences}
            \State $A_q \gets$ entire AI Overview text for query $q$
            \State $C_k \gets$ all chunks of website $k$
            \State Select $c^*_k = \arg\max_{c \in C_k}$ similarity$(c, A_q)$
        \Else 
            \State $snippet_k \gets$ search result snippet for website $k$ highlighted in conventional search
            \State Select $c^*_k \gets$ chunk containing $snippet_k$
        \EndIf
        \State Store representative chunk $c^*_k$ for website $k$
    \EndFor
\EndFor
\State \Return One representative chunk per website
\end{algorithmic}
\end{algorithm}

For the second type of analysis, we identify potential reference chunks at the sentence level within each query's AI summary. Specifically, for each sentence containing references, we determine one ``candidate'' chunk from each website that could potentially serve as supporting evidence for that sentence. This process involves iteratively examining all sentences with references and all the website chunks to identify the most relevant chunk for every (sentence, website) pair, as detailed in Algorithm \ref{alg:chunk_selection_rq1}.

\begin{algorithm}[h]
\caption{Chunk Selection: One Chunk per (AI Overview's sentence, Website) pair}
\label{alg:chunk_selection_rq1}
\begin{algorithmic}[1]
\Require AI Overview responses, Website content
\For{each query $q$}
   \State Retrieve all sentences $S_q$ that cite at least one website
   \State Retrieve all related websites $K_q$ from both AI Overview and search results
   \For{each sentence $s \in S_q$}
       \For{each website $k \in K_q$}
           \State Divide website $k$ into chunks $C_k$
           \State Identify the most relevant chunk: $\arg\max_{c \in C_k}$ similarity$(c, s)$
           \State Record mapping of query $q$, sentence $s$, website $k$, and segment $c$
       \EndFor
   \EndFor
\EndFor
\end{algorithmic}
\end{algorithm}
\par

Based on the aforementioned procedures, we construct two datasets. The first dataset compiles one chunk per website with binary indicators ($ChatCite$) of whether the AI Overview cites the website. The second dataset compiles one chunk for each (sentence, website) pair with binary indicators ($SentenceCite$) of whether the website is cited by the specific sentence. As outlined in Research Question 1, we construct variables based on these chunks' content to facilitate subsequent analyses. Specifically, we calculate perplexity for each chunk in both datasets and compute semantic similarity for every pair of chunks in the first dataset.

Perplexity measures a model's uncertainty when predicting text, with lower values indicating higher predictability from the model's perspective \citep{gonen2022demystifying}. We employ Google's Gemma-2B model to obtain the conditional probability of each token based on the preceding token sequence, then average across all tokens within each chunk: $PPL = \exp\left(-\frac{1}{N}\sum_{i=1}^{N}\log P(w_i|w_1,\ldots,w_{i-1})\right)$. For semantic similarity, we embed both chunks of a pair using the Universal Sentence Encoder and calculate pairwise cosine similarities between the resulting embedding pairs \citep{cer2018universal}.

\begin{table}[!h] \centering 
\TABLE{Summary Statistics for RQ1 Dataset\label{tab:summary_statistics_rq1}}
{
\begin{tabular}{@{\extracolsep{10pt}}lrrrrrr} 
\\[-1.8ex]\hline 
\hline \\[-1.8ex] 
Statistic & \multicolumn{1}{c}{N} & \multicolumn{1}{c}{Mean} & \multicolumn{1}{c}{St. Dev.} & \multicolumn{1}{c}{Min} & \multicolumn{1}{c}{Median} & \multicolumn{1}{c}{Max} \\ 
\hline \\[-1.8ex] 
\textbf{Website-level Data}   \\
\textit{ChatCite} & 98,477 & 0.47 & 0.50 & 0 & 0 & 1 \\ 
\textit{PPL} & 98,477 & 16.42 & 9.52 & 1.09 & 13.86 & 82 \\ 
\textit{Similarity} & 1,261,914 & 0.54 & 0.16 & -0.10 & 0.55 & 0.99 \\ 
\textbf{(Sentence-Website)-Level Data}   \\
\textit{SentenceCite} & 253,502 & 0.04 & 0.21 & 0 & 0 & 1 \\ 
\textit{PPL} & 253,502 & 16.95 & 10.59 & 1.21 & 14.25 & 197 \\ 
\hline \\[-1.8ex] 
\end{tabular} }
{
\footnotesize
\centering
Note: ``Similarity'' has more observations because we calculate the value for each pair from the same query.}
\end{table}

\subsubsection{Empirical Results for RQ1}\ 

Utilizing the aforementioned dataset derived from direct interaction with Google's search infrastructure, identification of relevant chunks, and construction of corresponding metrics, we now conduct a formal comparative analysis between chunks cited by Google's AI Overview and those excluded from citation.

Specifically, in the website-level dataset, for each query $q$, we extract one representative chunk $c$ from each website. Each chunk possesses a perplexity measure $PPL_{q,c}$ and may or may not be cited by the AI Overview, as indicated by the binary variable $ChatCite_{q,c}$. In the sentence-website level dataset, we consider that each AI Overview comprises $S_q$ sentences (indexed by $s=1,\ldots,S_q$), each of which may reference supporting sources. For every sentence $s$, we retrieve one candidate chunk $c$ from each website, characterized by perplexity $PPL_{q,s,c}$. We also record whether sentence $s$ cites the corresponding website, denoted by $SentenceCite_{q,s,c}$.

The perplexity measure assumes particular theoretical significance within the generative search engine paradigm, as it quantifies the predictability of textual content from a language model's perspective (i.e., low perplexity represents high predictability), a metric extensively employed in evaluating both textual characteristics and generative model performance. For instance, researchers conducting domain-specific fine-tuning of general-purpose language models (e.g., programming, biomedical, and legal domains) monitor perplexity during training procedures to assess model adaptation to specialized knowledge domains \citep{bolton2024biomedlm}. Furthermore, perplexity serves as a diagnostic indicator for identifying AI-generated text \citep{xu2024detecting}, reflecting one fundamental mechanism of auto-regressive models: these architectures sequentially select tokens based on predictability scores, with greedy decoding strategies favoring the most predictable tokens at each generation step, thereby systematically reducing overall text perplexity \citep{vaswani2017attention}.

Formally, we estimate regression models in which the binary citation outcomes \(ChatCite_{q,c}\) and \(SentenceCite_{q,s,c}\) serve as the dependent variables, and the chunk's perplexity \(PPL_{q,c}\) (at the sentence level, \(PPL_{q,s,c}\)) is the primary explanatory variable. Each specification includes query-level fixed effects \(Q_q\) to absorb unobserved, query-specific heterogeneity; consequently, the estimated association between perplexity and the probability of being cited is identified from within-query variation.
\begin{equation}
\label{eq:1}
    ChatCite_{q,c} = \beta_0 + \beta_1 * PPL_{q,c} + Q_q + \epsilon_{q,c}
\end{equation}
\begin{equation}
\label{eq:2}
    SentenceCite_{q,s,c} = \beta_0 + \beta_1 * PPL_{q,s,c} + Q_q + \epsilon_{q,s,c}
\end{equation}

We implement two distinct econometric specifications: Ordinary Least Squares (OLS) regression (constituting a linear probability model) and logistic regression. As demonstrated in Table \ref{tab:AI_perplex}, our empirical analysis reveals that chunks exhibiting lower perplexity demonstrate significantly higher citation probabilities within AI Overview responses. Specifically, a one standard deviation decrease in perplexity (9.52) corresponds to an increase in citation probability from an average of 47\% (shown in Table \ref{tab:summary_statistics_rq1}) to 56\%. This finding suggests that language models systematically exhibit preferences for content that aligns with their intrinsic generation patterns and stylistic characteristics. Analogous results emerge when examining sentence-level citations. Additionally, this perplexity effect is absent in conventional search ranking lists, as demonstrated in the Appendix \ref{appendix:ppl_conventional}. Our results further corroborate the theoretical mechanism previously articulated: when LLMs incorporate external source material into their outputs, the generated content inherits the perplexity characteristics of those sources. Consequently, when external sources exhibit high predictability from the LLM's perspective, their integration facilitates the model's operational objective of generating predictable content through sequential token-by-token processing.

\begin{table}[H]
\centering
\caption{AI Overview's Perplexity Effect}
\label{tab:AI_perplex}
\newcolumntype{L}[1]{>{\raggedright\arraybackslash}p{#1}}
\newcolumntype{C}[1]{>{\centering\arraybackslash}p{#1}}
\newcolumntype{R}[1]{>{\raggedleft\arraybackslash}p{#1}}
\renewcommand\arraystretch{0.5}
\begin{tabular}{L{2.5cm} c c c c}
\hline\hline
& \multicolumn{2}{c}{\(ChatCite\)} & \multicolumn{2}{c}{\(SentenceCite\)} \\
\cline{2-5}
& \(LPM\) & \(Logit\) & \(LPM\) & \(Logit\) \\
\cline{2-5}
& (1) & (2) & (3) & (4) \\
\hline
\(PPL\) 
  & -0.0098*** & -0.0480*** & -0.0015*** & -0.0581*** \\
& (0.0002) & (0.0009) & (0.0000) & (0.0015) \\
\hline
Query FE    & Yes & Yes & Yes & Yes \\
Observations     & 98,477 & 98,477 & 253,501 & 253,501 \\
\(R^2\) or \(Chi^2\)     & 0.13 & 3300.85 & 0.06  & 1820.31 \\
\hline\hline
\end{tabular}
\begin{tablenotes}
\footnotesize
\centerline{Note: Cluster-robust standard errors are reported at the query level. *** $p$ $<$ 0.01, ** $p$ $<$ 0.05, * $p$ $<$ 0.1.}
\end{tablenotes}
\end{table}

Furthermore, beyond examining the stylistic preferences of generative search engines, we investigate the semantic properties of their outputs, given their direct influence on user information exposure and subsequent behavioral responses. Extensive prior research has demonstrated that information retrieval systems critically influence opinion formation and belief construction processes through their information diversity characteristics \citep{epstein2015search, pariser2011filter}. Given that generative search engines constitute an emergent class of information retrieval systems, they similarly warrant scrutiny around the diversity metrics.

Specifically, using the website-level dataset, we compare semantic similarity among AI-cited chunks versus those highlighted by conventional search engines\footnote{We utilize the website-level rather than the sentence-website level dataset here, as our focus lies on the overall information diversity of AI Overview.}. For each pair of chunks (denoted by $c$ and $c'$), we calculate semantic similarity ($Similarity_{q,c,c'}$) using their respective embeddings. We focus on comparing chunk pairs where both chunks are cited by AI ($BothCite_{q,c,c'} = 1$) versus those that are not both cited by AI ($BothCite_{q,c,c'} = 0$). We consider two specifications for the control group ($BothCite_{q,c,c'} = 0$): one includes only pairs with both chunks non-cited, while the other additionally encompasses pairs with one cited and one non-cited chunk. Using these two distinct samples, we regress the similarity measure on the citation indicator, incorporating query-specific fixed effects $Q_q$.
\begin{equation}
\label{eq:3}
    Similarity_{q,c,c'} = \beta_0 + \beta_1 * BothCite_{q,c,c'} + Q_q + \epsilon_{q,c,c'}
\end{equation}

Regression results for both samples are presented in Table \ref{tab:AI_simi}. Our empirical analysis reveals that websites cited by generative search engines exhibit significantly higher semantic similarity on average compared to conventionally ranked results ($\beta_1 = 0.0365$ or $\beta_1 = 0.0492$, $p < 0.01$). This finding is particularly noteworthy given that the majority of cited websites do not appear within the top organic search and represent relatively niche sources that conventional search algorithms would not prioritize for query relevance. Nevertheless, these cited sources demonstrate substantially greater semantic similarity (low information diversity). Consequently, despite AI summaries typically presenting information through multiple sub-points and incorporating seemingly diverse niche sources, users encounter a narrowed range of semantic perspectives compared to conventional search results.

\begin{table}[H]
\centering
\caption{AI Overview's Semantic Similarity Effect}
\label{tab:AI_simi}
\newcolumntype{L}[1]{>{\raggedright\arraybackslash}p{#1}}
\newcolumntype{C}[1]{>{\centering\arraybackslash}p{#1}}
\newcolumntype{R}[1]{>{\raggedleft\arraybackslash}p{#1}}
\renewcommand\arraystretch{0.5}
\begin{tabular}{L{3cm} c c}
\hline\hline
& \multicolumn{2}{c}{\(Similarity\)} \\
\cline{2-3}
 & (1) & (2) \\
\hline
\(BothCite\) 
  & 0.0365*** & 0.0492*** \\
& (0.0020) & (0.0015) \\
\hline
Cross-Category     & No & Yes\\
Query FE    & Yes & Yes \\
Observations     & 674,086 & 1,261,914\\
\(R^2\)          & 0.31   & 0.31 \\
\hline\hline
\end{tabular}
\begin{tablenotes}
\footnotesize
\centering
Note: Cluster-robust standard errors are reported at the query level. *** $p$ $<$ 0.01, ** $p$ $<$ 0.05, * $p$ $<$ 0.1; \\ ``Cross-Category'' indicates whether $BothCite = 0$ includes pairs with one cited chunk and one non-cited chunk.
\end{tablenotes}
\end{table}

\subsection{Origins of Citation Criteria in Generative Search Engines (RQ2)}
\label{sec: RAG_criteria}

\subsubsection{RAG Experimental Design and Variables}\

In our exploration of RQ1, we have documented the stylistic and semantic preferences exhibited by Google's AI Overview. However, given the additional engineering processes required to deploy the underlying LLM within generative search engines, these preferences may be specifically and artificially designed for such systems. While these preferences could theoretically emerge naturally, we seek to test their boundaries through direct interaction with LLM-based RAG systems.

To do so, we leverage the website-level dataset constructed in RQ1 as the same source, but use Google's Gemini RAG API to provide answers to each query with the same query, as summarized by Algorithm \ref{alg:rag_complete}. Specifically, for each query, we combine this query's related chunks from the website-level dataset into one PDF as the source file, each chunk is labeled with an index. Since this is in a controlled environment, we can safely avoid the confounding effect from position bias by randomly ranking all these chunks. Then, we call the Gemini RAG API and ask the AI to mimic Google AI Overview's function and answer the query, and it can cite chunks from the source file with specific indexes to support the answer\footnote{The specific prompt can be found in Appendix \ref{appendix: prompt}.}. Therefore, for each query, we can know exactly which chunks the RAG system chooses to cite, and we relabel the original dataset with $RAGCite_{q,c}$. Since the original sources stay the same, the $PPL_{q,c}$ and $Similarity_{q,c,c'}$ will be the same. The summary statistics are again shown in Table \ref{tab:summary_statistics_rq2} for references.

To address this objective, we employ the website-level dataset constructed in RQ1, utilizing Google's Gemini RAG API to generate responses for each query, as delineated in Algorithm~\ref{alg:rag_complete}. Specifically, for each query $q$, we aggregate the corresponding chunks from the website-level dataset into a single PDF document that serves as the source file, with each chunk assigned a unique index. This controlled experimental design enables us to mitigate potential confounding effects arising from position bias through random ordering of all chunks \citep{shi2024judging}. Subsequently, we invoke the Gemini RAG API with instructions to emulate the functionality of Google AI Overview in answering the query, whereby the system can reference specific chunks from the source file using their corresponding indices to substantiate its response. We precisely identify which chunks the RAG system selects for citation for each query, thereby relabeling the original dataset with $RAGCite_{q,c}$. Given that the source materials remain constant, the values of $\mathit{PPL}_{q,c}$ and $\mathit{Similarity}_{q,c,c'}$ remain unchanged. The summary statistics for the new dataset are presented in Table~\ref{tab:summary_statistics_rq2} for reference.

\begin{algorithm}[h]
\caption{RAG Experimental Protocol for RQ2}
\label{alg:rag_complete}
\begin{algorithmic}[1]
\Require Query–website mappings from RQ1
\State Initialize dataset for RQ2 analysis
\For{each query $i$}
   \State Compile a labeled document of relevant website chunks
   \State Submit the document to the Gemini RAG system together with the query
   \State Prompt the model to provide an integrated summary and to indicate citations
   \State Record the cited sources
\EndFor
\end{algorithmic}
\end{algorithm}

\begin{table}[!h] \centering 
\TABLE{Summary Statistics for RQ2 Dataset\label{tab:summary_statistics_rq2}}
{
\begin{tabular}{@{\extracolsep{10pt}}lrrrrrr} 
\\[-1.8ex]\hline 
\hline \\[-1.8ex] 
Statistic & \multicolumn{1}{c}{N} & \multicolumn{1}{c}{Mean} & \multicolumn{1}{c}{St. Dev.} & \multicolumn{1}{c}{Min} & \multicolumn{1}{c}{Median} & \multicolumn{1}{c}{Max} \\ 
\hline \\[-1.8ex] 
\textit{RAGCite} & 98,477 & 0.33 & 0.47 & 0 & 0 & 1 \\ 
\textit{PPL} & 98,477 & 16.42 & 9.52 & 1.09 & 13.86 & 82 \\ 
\textit{Similarity} & 1,261,914 & 0.54 & 0.16 & -0.10 & 0.55 & 0.99 \\ 
\hline \\[-1.8ex] 
\end{tabular} }
{}
\end{table} 

\subsubsection{Empirical Results for RQ2} \

Employing the previously described dataset, we examine whether the observed stylistic and semantic preferences represent intrinsic model characteristics or constitute engineered modifications specifically designed for generative search functionality. Our experimental framework maintains identical variable construction methodologies, with the exception that citation labels are now generated through Gemini's RAG API rather than Google's AI Overview system.

In the revised dataset, we define the citation outcome variable as $RAGCite_{q,c}$, which takes the value of 1 if chunk $c$ receives a citation in the RAG response to query $q$, and 0 otherwise. Each chunk's perplexity is denoted as $PPL_{c}$, while its positional placement within the source document is represented by $Pos_{c}$. After controlling for query-level fixed effects $Q_q$, we specify the following empirical model to examine the perplexity effect within the RAG system:
\begin{equation}
\label{eq:4}
    RAGCite_{q,c} = \beta_0 + \beta_1 * PPL_{q,c} + \beta_2 * Pos_{q,c} + Q_q + \epsilon_{q,c}
\end{equation}

Additionally, utilizing the revised citation labels, we maintain identical semantic similarity construction procedures, and employ a specification analogous to that presented in the preceding section to evaluate differences in semantic similarity among cited chunks relative to that among non-cited chunks. The sole modification is the citation determination mechanism, which is now governed by the RAG system. Consequently, the independent variable of interest transitions from $BothCite_{q,c,c'}$ to $BothRAGCite_{q,c,c'}$:
\begin{equation}
\label{eq:5}
    Similarity_{q,c,c'} = \beta_0 + \beta_1 * BothRAGCite_{q,c,c'} + Q_q + \epsilon_{q,c,c'}
\end{equation}

Interestingly, our regression analyses demonstrate that both observed preferences manifest consistently within the RAG system architecture. First, the Gemini-based RAG exhibits a significant propensity to select content with lower perplexity (Table \ref{tab:rag_perplex}). Second, chunks cited by the Gemini-based RAG demonstrate significantly greater semantic similarity among themselves relative to the semantic similarity observed among non-cited chunks (Table \ref{tab:rag_simi}). This consistency provides compelling evidence that these mechanisms are deeply rooted in the underlying language models: they preferentially cite content that is predictable from their computational perspective, thereby aligning with the objective function of such models, and select semantically similar content to ensure coherence in their generated outputs.

\begin{table}[h]
\centering
\caption{RAG's Perplexity Effect and Position Bias}
\label{tab:rag_perplex}
\newcolumntype{L}[1]{>{\raggedright\arraybackslash}p{#1}}
\newcolumntype{C}[1]{>{\centering\arraybackslash}p{#1}}
\newcolumntype{R}[1]{>{\raggedleft\arraybackslash}p{#1}}
\renewcommand\arraystretch{0.5}
\begin{tabular}{L{2.5cm} c c c c}
\hline\hline
& \multicolumn{2}{c}{\(RAGCite\)}\\
\cline{2-3}
& \(LPM\) & \(Logit\) \\
\cline{2-3}
& (1) & (2)\\
\hline
\(PPL\) 
  & -0.0067*** & -0.0427*** \\
& (0.0002) & (0.0010)  \\
\(Pos\) 
  & -0.0145*** & -0.0849*** \\
& (0.0003) & (0.0011)  \\
\hline
Query FE    & Yes & Yes \\
Observations     & 98,477 & 98,477  \\
\(R^2\) or \(Chi^2\)    & 0.21   & 10,248\\
\hline\hline
\end{tabular}
\begin{tablenotes}
\footnotesize
\centerline{Note: Cluster-robust standard errors are reported at the query level. *** $p$ $<$ 0.01, ** $p$ $<$ 0.05, * $p$ $<$ 0.1.}
\end{tablenotes}
\end{table}

\begin{table}[h]
\centering
\caption{RAG's Semantic Similarity Effect}
\label{tab:rag_simi}
\newcolumntype{L}[1]{>{\raggedright\arraybackslash}p{#1}}
\newcolumntype{C}[1]{>{\centering\arraybackslash}p{#1}}
\newcolumntype{R}[1]{>{\raggedleft\arraybackslash}p{#1}}
\renewcommand\arraystretch{0.5}
\begin{tabular}{L{3cm} c c}
\hline\hline
& \multicolumn{2}{c}{\(Similarity\)} \\
\cline{2-3}
 & (1) & (2) \\
\hline
\(BothRAGCite\) 
  & 0.1206***  & 0.0964*** \\
& (0.0016) & (0.0013) \\
\hline
Cross-Category     & No & Yes\\
Query FE    & Yes & Yes \\
Observations     & 766,243 & 1,261,914\\
\(R^2\)          & 0.36   & 0.32 \\
\hline\hline
\end{tabular}
\begin{tablenotes}
\footnotesize
\centerline{Note: Cluster-robust standard errors are reported at the query level. *** $p$ $<$ 0.01, ** $p$ $<$ 0.05, * $p$ $<$ 0.1.}
\end{tablenotes}
\end{table}

Furthermore, our analysis reveals that positional placement within the source document significantly influences the probability of selection for citation ($\beta_2$ of Table \ref{tab:rag_perplex}). Given that these systems exhibit a preference for content positioned at the document's beginning, this finding suggests an additional content optimization strategy for website proprietors: they may strategically place core arguments and key information at the beginning of websites to ensure that primary content remains ``visible'' to generative retrieval systems and receives favorable consideration for citation inclusion.

\subsection{Impact of LLM-Driven Polishing on AI summary's Content Diversity (RQ3)}
\label{sec: LLM_polish}

\subsubsection{LLM-driven Polishing, RAG, and Variables}\

Having explored the preferences of generative search engines and verified the underlying mechanisms through RAG API analysis, we now examine the implications from website owners' perspectives. A natural strategy for increasing content predictability from an LLM's computational perspective is to employ LLMs for content refinement, as such polished content, being generated token-by-token through the LLM's conditional probability optimization process, typically exhibits lower perplexity. Moreover, utilizing LLMs for marketing content generation has already emerged as a prevalent application \citep{ye2025lola, reisenbichler2022frontiers}. However, such refinement practices may inadvertently result in content homogenization across different website providers, thereby reducing the diversity of the information ecosystem. To assess how LLM-based content polishing influences information diversity in generative search, we conduct a controlled experiment using Gemini's API, simulating a strategic environment where AI systems from different parties interact competitively.

Prior to processing source files through the RAG API as described in Section \ref{sec: RAG_criteria}, we implement two types of content optimization using Google's Gemini 1.5 Flash model for each chunk. The first approach employs a general refinement procedure that enhances clarity and engagement while preserving substantive meaning. This optimization simulates scenarios where website owners utilize LLMs to assist in the content creation process. The second approach implements goal-oriented optimization that explicitly structures content to increase the likelihood of citation in Google's AI Overview. This mimics the strategic behavior whereby website owners deliberately consider generative search engine exposure during content development. Both prompt specifications are provided in Appendix \ref{appendix: prompt}.

Following this optimization phase, we repeat the RAG experiments developed in RQ2 under three parallel conditions: using the original content, using the polished version, and using the AI-optimized version. 
\begin{algorithm}[h]
\caption{LLM-driven Content Optimization and RAG Testing}
\label{alg:llm_optimization}
\begin{algorithmic}[1]
\Require Original chunks from RQ1
\State \textbf{Phase 1: Content Optimization}
\For{each chunk $c$ in dataset}
    \State Produce general polished version ($c_{pol}$)

    \State Produce AI-optimized version($c_{opt}$)
    \State Store: $(c, c_{pol}, c_{opt})$
\EndFor
\State \textbf{Phase 2: RAG Experiments (3 conditions)}
\For{each condition $\in$ \{original, polished, AI-optimized\}}
    \State Create PDFs using chunks from current condition
    \State Execute RAG protocol (same as RQ2)
    \State Record citation results for each condition
\EndFor
\State \Return Citation results for all three conditions
\end{algorithmic}
\end{algorithm}

Implementing the same procedure outlined in Section 4.2, we obtain the citation list for each query and calculate the pairwise similarity between all cited chunks, enabling us to measure and compare information diversity across three experimental conditions. Further, we compute the total number of citations and the perplexity of the AI-generated summary for each query to provide additional comparative analysis. The summary statistics are provided in Table \ref{tab:summary_statistics_rq3}\footnote{With 4,060 queries and three conditions, we obtain 12,180 observations for query-level measurements. For similarity analyses, we generate 1,261,914 pairs under each condition, yielding 3,785,742 total observations.}.

\begin{table}[!h] \centering 
\TABLE{Summary Statistics for RQ3 Dataset\label{tab:summary_statistics_rq3}}
{
\begin{tabular}{@{\extracolsep{10pt}}lrrrrrr} 
\\[-1.8ex]\hline 
\hline \\[-1.8ex] 
Statistic & \multicolumn{1}{c}{N} & \multicolumn{1}{c}{Mean} & \multicolumn{1}{c}{St. Dev.} & \multicolumn{1}{c}{Min} & \multicolumn{1}{c}{Median} & \multicolumn{1}{c}{Max} \\ 
\hline \\[-1.8ex] 
\textit{Similarity} & 3,785,742 & 0.53 & 0.16 & -0.15 & 0.54 & 0.99 \\ 
\textit{NumCite} & 12,180 & 9.02 & 5.08 & 1 & 8 & 27 \\ 
\textit{OutputPPL} & 12,180 & 8.37 & 8.81 & 2.72 & 7.88 & 95 \\ 
\hline \\[-1.8ex] 
\end{tabular} }
{}
\end{table}

\subsubsection{Empirical Results for RQ3}\ 

Utilizing the three distinct content treatment conditions (original chunks, AI-polished chunks, and objective-oriented AI-polished chunks) and their corresponding RAG selection outcomes, we examine the counterfactual effects of LLM-based content optimization prior to RAG processing. This analysis simulates real-world scenarios wherein website proprietors deploy LLMs for content generation and optimization purposes.

Initially, we assess overall semantic similarity patterns across all chunks under the three experimental conditions. We denote the original content condition as $T = 0$, AI-polished content without specific objectives as $T = 1$, and AI-polished content with citation optimization objectives as $T = 2$. For each pair of chunks ($c$ and $c'$) derived from query $q$ under treatment condition $T$, we calculate the pairwise semantic similarity measure $Similarity_{T,q,c,c'}$. We regress the similarity on the condition indicator, with query-specific fixed effects included as well. Additionally, we conduct a focused analysis restricting the sample to cited websites exclusively, examining how semantic similarity among cited sources varies across different input treatment conditions. Both analyses are presented in Table \ref{tab:polish_simi}.
\begin{equation}
\label{eq:6}
    Similarity_{T,q,c,c'} = \beta_0 + \beta_1 \cdot \mathbf{1}(T = 1)_{T,q,c,c'} + \beta_2 \cdot \mathbf{1}(T = 2)_{T,q,c,c'} + Q_q + \epsilon_{T,q,c,c'}
\end{equation}

\begin{table}[h]
\centering
\caption{Impact of Sources' LLM Polishing on (Cited) Content's Similarity}
\label{tab:polish_simi}
\newcolumntype{L}[1]{>{\raggedright\arraybackslash}p{#1}}
\newcolumntype{C}[1]{>{\centering\arraybackslash}p{#1}}
\newcolumntype{R}[1]{>{\raggedleft\arraybackslash}p{#1}}
\renewcommand\arraystretch{0.5}
\begin{tabular}{L{3cm} c c}
\hline\hline
& \multicolumn{2}{c}{\(Similarity\)} \\
\cline{2-3}
 & (1) & (2) \\
\hline
\(T = 1\) 
  & -0.0096*** & -0.0319*** \\
& (0.0004) & (0.0011) \\
\(T = 2\) 
  & -0.0091*** & -0.0722*** \\
& (0.0005) & (0.0011) \\
\hline
Sample     & All & Cited Only\\
Query FE    & Yes & Yes \\
Observations     & 3,785,742 & 597,533\\
\(R^2\)          & 0.29   & 0.42 \\
\hline\hline
\end{tabular}
\begin{tablenotes}
\footnotesize
\centerline{Note: Cluster-robust standard errors are reported at the query level. *** $p$ $<$ 0.01, ** $p$ $<$ 0.05, * $p$ $<$ 0.1.}
\end{tablenotes}
\end{table}

Our empirical findings reveal a surprising pattern: while overall semantic diversity decreases at a small magnitude under AI polishing ($\beta_1 = -0.0096$, Column (1), Table \ref{tab:polish_simi}), semantic similarity among cited websites decreases significantly ($\beta_1 = -0.0319$, Column (2), Table \ref{tab:polish_simi}). This phenomenon likely occurs because such content polishing preserves original semantic content while rendering previously unpredictable material sufficiently predictable for LLM consideration. Consequently, as the pool of stylistically compatible content (characterized by low perplexity) expands, the RAG system can incorporate a broader range of external sources, thereby enhancing citation diversity among selected websites. Moreover, objective-oriented polishing yields even more substantial improvements in outcome diversity ($\beta_2 = -0.0722$, Column (2), Table \ref{tab:polish_simi}). This result indicates that despite the absence of detailed instructions, LLMs demonstrate inherent understanding of enhancement strategies, enabling the RAG system to consider an expanded set of potential sources autonomously.

We further investigate this underlying mechanism through two complementary analyses. First, we examine the number of chunks cited per query $NumCite_{T,q}$ as a function of treatment conditions. Second, we analyze the perplexity of RAG-generated responses $OutputPPL_{T,q}$ under different treatment scenarios. Both dependent variables are modeled using the following specification:
\begin{equation}
\label{eq:7}
    Outcome_{T,q} = \beta_0 + \beta_1 \cdot \mathbf{1}(T = 1)_{T,q} + \beta_2 \cdot \mathbf{1}(T = 2)_{T,q} + \epsilon_{T,q}
\end{equation}

\begin{table}[h]
\centering
\caption{Impact of Sources' LLM Polishing on Outputs' Attributes}
\label{tab:polish_output}
\newcolumntype{L}[1]{>{\raggedright\arraybackslash}p{#1}}
\newcolumntype{C}[1]{>{\centering\arraybackslash}p{#1}}
\newcolumntype{R}[1]{>{\raggedleft\arraybackslash}p{#1}}
\renewcommand\arraystretch{0.5}
\begin{tabular}{L{3cm} c c}
\hline\hline
& \(NumCite\) & \(OutputPPL\) \\
\cline{2-3}
 & (1) & (2) \\
\hline
\(T = 1\) 
  & 1.0744*** & -0.0431 \\
& (0.1071) & (0.0823) \\
\(T = 2\) 
  & 2.1059*** & -0.0451 \\
& (0.1098) & (0.0762) \\
\hline
Observations     & 12,180  & 12,180 \\
\(R^2\)          & 0.03   & 0.02 \\
\hline\hline
\end{tabular}
\begin{tablenotes}
\footnotesize
\centerline{Note: Robust standard errors are reported. *** $p$ $<$ 0.01, ** $p$ $<$ 0.05, * $p$ $<$ 0.1.}
\end{tablenotes}
\end{table}

As demonstrated in Column (1) of Table \ref{tab:polish_output}, we observe a significant increase in the number of citations following content polishing, with the effect being particularly pronounced when optimization includes the citation objective. Importantly, this expanded citation scope does not compromise the predictability of RAG outputs; conversely, both treatment conditions yield marginally decreased output perplexity (Column (2), Table \ref{tab:polish_output}). These collective findings indicate that LLM-based content polishing does not directly induce homogenization effects. Rather, by enhancing content accessibility and interpretability for language models, such optimization strategies may actually improve the information diversity available to users through generative retrieval responses.

\section{User-Side Consequences of LLM-based polishing (RQ4)}

Our prior analyses demonstrate that LLM-driven content polishing, while optimized for citation probability in generative search engines, paradoxically enhances information diversity within AI-generated summaries. This finding naturally leads to questions regarding downstream implications: How does this transformation of the information ecosystem influence end users' information acquisition experiences? To investigate the effects, we develop a controlled experimental platform replicating Google's AI Overview functionality and recruit participants from Prolific to conduct a between-subjects randomized controlled trial addressing our final research question (RQ4).

\subsection{Experimental Design}  \

We design and implement a custom generative search platform that replicates the functionality of Google AI Overview while enabling comprehensive capture of user behavioral data within our controlled environment (see Appendix \ref{appendix: experiment} for the interface). Utilizing this platform, we recruit 150 participants through Prolific to conduct a between-subjects randomized experiment, thereby investigating the effects of LLM polishing on website source evaluation and utilization.

To reflect what users obtain from generative search engines, we ask participants to complete a writing task \citep{noy2023experimental}. Since we seek to polish websites in the treatment condition, we focus on one specific topic so that polished website materials can be prepared in advance. We select the teenagers' smartphone ban policy as our research topic: participants assume the role of education policy consultants, tasked with researching and recommending whether a public school district should implement a smartphone ban during school hours. We choose this task for several reasons: (1) it represents a contemporary policy debate with no definitive correct answer, (2) it requires synthesizing diverse perspectives, including educational, psychological, and practical considerations, and (3) it mirrors the type of complex decision-making processes that generative search engines are designed to support. The specific instructions that participants receive are also presented in Appendix \ref{appendix: experiment}.

To compile website sources for user searches, we conduct searches on topics encompassing the advantages and disadvantages of smartphone bans, potential enforcement mechanisms, alternative approaches to address smartphone usage limitations, and related case studies. We obtain relevant website links from Google (including AI Overview) and Bing (including Microsoft Copilot), employing the following extraction methodology: for generative search results, we extract all referenced websites, while for conventional search results, we extract websites from the first page of organic results. Across all search channels and keywords, we compile 65 unique websites pertaining to this topic. We then download and process the complete content of these 65 websites to establish our baseline information source (Source ``Control''). For the treatment condition, we apply the identical LLM-polishing procedure with the citation optimization objective described in Section \ref{sec: LLM_polish} to generate a parallel set of polished content (Source ``Treatment''). These two source conditions are the sole experimental manipulation that participants encounter during the study.

Based on these sources, we construct a custom web-based generative search engine utilizing the Gemini RAG API to conduct the randomized experiment, with detailed procedural information provided in Appendix \ref{appendix: experiment}. When users enter the platform and provide their Prolific ID, the underlying treatment assignment becomes fixed, with 50\% allocated to the treatment group and 50\% to the control group, respectively. Users then review the instructions and query our generative search engine, which returns AI-generated summaries with relevant website citations, replicating Google AI Overview's functionality. Both experimental groups encounter an identical user interface, with the sole distinction being the previously mentioned source manipulation: the RAG system generates responses for treatment group participants based on Source ``Treatment,'' while responses for control group participants are based on Source ``Control.'' Users can conduct unlimited searches, with each query-response interaction recorded for subsequent analysis. To ensure task engagement and prevent mechanical reproduction of content, we disable paste functionality in both query and response fields, requiring participants to actively synthesize information using their own formulation.

Each participant submits one final recommendation (100-300 words), after which the platform redirects participants to a post-task survey, with the complete questionnaire presented in Appendix \ref{appendix: survey}. We also collect standard demographic information to conduct randomization checks and explore potential treatment effect heterogeneities. Three participants fail to finish the survey and are thus dropped from the analyses.

\subsection{Experimental Data Analyses and Results}

Before formally testing the treatment effect, we validate the randomization assumption by conducting pairwise t-tests and Kolmogorov-Smirnov (KS) tests on all demographic variables and prior experience with relevant tools. As demonstrated in Appendix \ref{appendix: randomization}, we confirm that participants in the treatment and control groups exhibit no statistically significant differences across these dimensions.

Based on the dataset collected from the experiment, we directly compare the submitted recommendations and task efficiency across the two experimental groups. Motivated by our exploration of RQ3, where such treatments increase the information diversity of AI summary outputs, we capture the information diversity of participants' submissions using the Vendi score \citep{friedman2022vendi}. Regarding task efficiency, we calculate the time spent by each participant. Formally, for each participant $i$, we derive $Information_i$ (in Vendi score) and $TimeSpent_i$ (in minutes) as two primary outcomes of interest ($Outcome_i$). Each participant possesses a treatment status $Treat_i$, and we estimate the following regression models to derive the treatment effects.
\begin{equation}
\label{eq:8}
    Outcome_{i} = \beta_0 + \beta_1 \cdot Treat_i + \epsilon_{i}
\end{equation}

Furthermore, as education levels are commonly associated with individuals' adoption and usage of technologies \citep{ma2024learning, xu2023chatgpt}, we explore the heterogeneous treatment effects across different education levels. Specifically, we denote a participant's education level with a binary variable $College_i$, which equals 1 if the participant possesses a college degree or higher, and 0 otherwise. We then partition the participants into two subsamples and conduct subsample analyses using the same specification as Equation (\ref{eq:8}).

As demonstrated in Table \ref{tab:exp}, we find that treated participants tend to provide written submissions with significantly higher information diversity, an effect primarily driven by participants without college degrees. Meanwhile, participants with college degrees or higher tend to provide submissions with similar information diversity across both groups, but treated participants can complete the task in significantly less time.

These results reveal interesting behavioral patterns: relatively more educated individuals already possess the tendency to seek diverse information sources; LLM-polishing and the consequent information diversity enhancement enable them to increase their efficiency. However, less educated participants tend to rely on immediately available information, and their written expressions are more substantially influenced by the information enhancement per query.

\begin{table}[!h]
\centering
\caption{Experimental Results}
\label{tab:exp}
\newcolumntype{L}[1]{>{\raggedright\arraybackslash}p{#1}}
\newcolumntype{C}[1]{>{\centering\arraybackslash}p{#1}}
\newcolumntype{R}[1]{>{\raggedleft\arraybackslash}p{#1}}
\renewcommand\arraystretch{0.5}
\begin{tabular}{L{2.5cm} c c c c c c}
\hline\hline
& \multicolumn{3}{c}{\(Information\)} & \multicolumn{3}{c}{\(TimeSpent\)} \\
\cline{2-7}

& (1) & (2) & (3) & (4) & (5) & (6) \\
\hline
\(Treat\) 
  & 0.3554*** & 0.0690 & 0.6123*** & -1.5942***  & -2.9186*** & -0.3858 \\
& (0.0688) & (0.0873) & (0.0862) & (0.3033) & (0.3685) & (0.3303) \\
\hline
College    & All & 1 & 0 & All & 1 & 0 \\
Observations     & 147 & 68 & 79 & 147  & 68 & 79 \\
\(R^2\)          & 0.15   & 0.01  & 0.40 & 0.16  & 0.49   & 0.02  \\
\hline\hline
\end{tabular}
\begin{tablenotes}
\footnotesize
\centerline{Note: Robust standard errors are reported. *** $p$ $<$ 0.01, ** $p$ $<$ 0.05, * $p$ $<$ 0.1.}
\end{tablenotes}
\end{table}

\section{Robustness Checks}

We conduct a series of robustness checks, including a replication study, alternative samples, and alternative specifications. Due to space constraints, we summarize them in Appendix \ref{appendix: robust}.

\section{Discussion and Conclusion}

This concluding section synthesizes our empirical findings, delineates the theoretical contributions and practical implications of our research, and identifies several limitations alongside promising avenues for future scholarly investigation.

\subsection{Summary of Results}

Through systematic interactions with both generative and traditional search engines, complemented by controlled experimentation with human participants, we have systematically addressed each research question and generated several key empirical findings. First, our analysis reveals that generative engines exhibit systematic preferences for content characterized by high predictability (low perplexity) and semantic homogeneity among cited sources (RQ1). Second, we demonstrate that this citation behavior stems from fundamental characteristics embedded within the underlying large language model architecture rather than platform-specific engineering decisions (RQ2). Third, contrary to expectations of AI-induced homogenization effects, comprehensive content polishing through LLMs actually enhances information diversity within AI-generated summaries. This unexpected outcome results from the perplexity-reduction effects of LLM polishing, which expands the pool of content eligible for citation consideration by generative search systems (RQ3). Finally, our randomized controlled trial demonstrates that LLM-enhanced information diversity translates into measurable improvements in task efficiency and information acquisition among end users. Notably, we observe differential treatment effects across educational backgrounds: highly educated participants primarily benefit through enhanced efficiency, while participants with lower educational attainment experience gains in information comprehension and utilization (RQ4).

\subsection{Contributions and Implications}

Our research advances theoretical understanding across multiple domains. First, we contribute to the search engine optimization literature by identifying novel content provision criteria specific to generative search engines. We reveal citation preferences that were previously unobservable within traditional ranking algorithms, and more importantly, demonstrate that these preferences emerge from the intrinsic language model rather than explicit engineering design choices. Second, within the broader search engine ecosystem framework, we establish that LLM-based content polishing generates ``win-win'' outcomes: expanded website appearances within AI summaries while simultaneously enhancing the semantic diversity of user-facing content. Third, our behavioral experiment contributes to the literature on AI summaries by demonstrating that information increases in AI-generated summaries yield differential benefits across users with varying backgrounds in terms of both task efficiency and output information value, thereby advancing our understanding of human interactions with AI summary systems.

Our findings extend beyond academic interest to provide actionable insights for diverse stakeholder communities. First, for website proprietors and generative search engine optimization service providers, the connection between foundational language models and LLM-based AI Overview systems presents novel optimization strategies. Practitioners can leverage RAG systems built upon identical model families (e.g., Gemini) to conduct offline optimization testing prior to website deployment, or even utilize open-source variants (e.g., Gemma) to expedite the optimization process while maintaining cost efficiency. Second, for information-seeking users, awareness of AI Overview's inherent constraints is crucial: the system's design objective of generating coherent paragraph-form responses may result in systematically narrowed perspectives. While users benefit from enhanced search efficiency, comprehensive information gathering may still require supplementary search activities to achieve a holistic understanding. Third, our results demonstrate that LLM-based content polishing can yield beneficial outcomes under certain conditions. However, the intrinsic relationship between underlying language models and AI Overview systems creates potential vulnerabilities to strategic manipulation. Search engine operators may need to carefully engineer distinctions between RAG APIs and AI Overview functionalities, and simultaneously reconsider open-source model distribution strategies to mitigate potentially harmful gaming behaviors by website proprietors.

\subsection{Limitations and Future Directions}

Our study acknowledges several limitations that suggest promising directions for future research. First, while Google dominates the search engine market, the external validity of our findings may remain constrained. To address this limitation, we conducted robustness checks using Microsoft's Bing and examined underlying RAG systems to elucidate deeper mechanisms. Nevertheless, future research investigating how alternative generative search engine architectures might mitigate these tendencies would help establish the boundary conditions of our results. Second, regarding temporal validity, our findings may not persist indefinitely as these technologies evolve. However, our data collection spans multiple time periods (Bing data from 2023 and Google Overview data from 2025), demonstrating that the observed effects have persisted for approximately two and a half years since the initial deployment of generative search engines. Given the apparent permanence of generative search technologies, our investigation falls into the ``Big Questions, Half Answers'' category of IS research \citep{hosanagar2017senior}, and we hope it can stimulate further scholarly discourse. Finally, concerning human-AI interaction analysis, we acknowledge the inherent limitations of our online experimentation approach. Field studies utilizing proprietary platform data could yield more comprehensive insights into the broader behavioral impacts of AI Overview systems.

\bibliography{thebibliography}

\bibliographystyle{informs2014}

\newpage

\begin{APPENDICES}

\renewcommand{\thesection}{\Alph{section}}
\renewcommand{\thesubsection}{\Alph{section}.\arabic{subsection}}
\counterwithin{table}{section}
\counterwithin{figure}{section}
\counterwithin{algorithm}{section}
\renewcommand{\thealgorithm}{\Alph{section}\arabic{algorithm}}
\renewcommand{\thetable}{\Alph{section}\arabic{table}}
\renewcommand{\thefigure}{\Alph{section}\arabic{figure}}

\section{Summary of Robustness checks}
\label{appendix: robust}

Although Google dominates the search engine market, these mechanisms should be generalizable to different LLM-based generative search engines. Therefore, we utilize a dataset collected through interactions with New Bing (subsequently renamed Microsoft Copilot) in April 2023 and replicate our analyses. The detailed results are presented in Appendix \ref{appendix: bing}.

Furthermore, our analyses identify outlier websites with excessively high perplexity values that may influence the results; consequently, we repeat the analyses with these outliers excluded. The findings are documented in Appendix \ref{appendix: outlier}.

Additionally, we observe that conventional search engines yield slightly more websites on average, which may contribute to the higher similarity observed among generative search engines' cited websites. To address this potential confound, we restrict both search channels to the same number of websites for each query and rerun the similarity analyses, with results presented in Appendix \ref{appendix: samples}.

Finally, we evaluate alternative model specifications to assess sensitivity to different parametric forms, with findings reported in Appendix \ref{appendix: specification}.

\newpage
\section{Raw Data Collection Process}

\label{appendix:API}

In this appendix, we detail the procedures used for data collection and dataset construction. Algorithm \ref{alg:data_collection} illustrates our data collection process. 

\begin{algorithm}[h]
\caption{Data Collection Process}
\label{alg:data_collection}
\begin{algorithmic}[1]
\Require Query list $Q$ from HC3 corpus

\For{each query $q \in Q$}
    \State Initiate candidate website set $S_q$ 
    \State Send the query $q$ to Google Search via the SerpAPI general endpoint
    \If{the response indicates a request to AI Overview specific endpoint is required}
        \State Send the query $q$ to AI Overview specific endpoint
    \EndIf
    \State Extract AI Overview response text $t_q$
    \State $S_q \gets$  websites cited in AI Overview 
    \State $S_q \gets$  first page of organic search results
    \For {each website $i$ in $S_q$}
        \State Render the web page in a browser and extract its content $c_{qi}$
    \EndFor

\EndFor
\end{algorithmic}
\end{algorithm}

The unified collection process ensures that both AI Overview data and conventional search results are fully extracted for each query in the HC3 dataset. For each query from the HC3 dataset, we utilize the SerpAPI to obtain both the AI Overview results and the first page of organic results. 
Sometimes the general endpoint does not directly return Google's AI Overview but returns a parameter that indicates an additional request to the AI Overview specific endpoint is required\footnote{See \url{https://serpapi.com/blog/scrape-google-ai-overviews/}}. In that case, we would send another request accordingly. For the AI Overview part, we save both the response text and the URLs of the cited websites. 
For the conventional search results, we record the ranking position, title, URL, and snippet of each listed website. For each query $q$, we construct a candidate website set consisting of sites that appear either in the AI Overview or among the top conventional search results.

In the next step, we collect content from all candidate websites. We employ browser automation to ensure each web page is rendered dynamically, allowing us to capture the full content. We then filter out non-textual elements such as images and stylesheets, retaining only the textual content relevant for downstream analyses.

\newpage

\section{Prompt Content}
\label{appendix: prompt}

\textbf{System Prompt for RAG API:} \textit{Assume that you are the Google AI Overview generator, a feature integrated into Google Search that provides AI-generated summaries of search results. Please answer the following query based on the website content contained in the attached PDF file. Within the PDF file, there is a list of numbered paragraphs, each of which represents a website's content indicated by a unique ID in the format ``Source 11,'' etc. Please mimic Google AI Overview's answering style. For each sentence, if you can find references from the PDF, cite the specific ID of that website's content. For citations, use the EXACT format: \%\%\%X,Y,Z\%\%\%. Separate multiple source IDs with commas. Do NOT use any other citation format, such as (Source X). Example: ``This is an example statement. \%\%\%1,5,12\%\%\%.''}

\vspace{0.25in}

\noindent \textbf{System Prompt for Content Polishing:} \textit{Here is an excerpt from a webpage: '{excerpt}'. Please polish the excerpt so that it is clearer and more engaging. Try to keep the length roughly unchanged. Only return the polished excerpt itself.}

\vspace{0.25in}

\noindent \textbf{System Prompt for Content Polishing with the objective:} \textit{Here is an excerpt from a webpage: '{excerpt}'. Please polish the excerpt so that it is clearer and more engaging. Try to keep the length roughly unchanged. The primary goal is to make this specific excerpt (and, by extension, the overall webpage) more likely to be selected and highlighted by Google Search's AI Overview feature. Only return the polished excerpt itself.}

\newpage

\section{New Bing (Microsoft Copilot) Study}
\label{appendix: bing}

To ensure the robustness of our findings across different large language model architectures and search engine platforms, we replicate all the analysis on Microsoft New Bing, one of the earliest platforms that launched a generative search engine in February 2023\footnote{See \url{https://www.microsoft.com/en-us/edge/features/the-new-bing}.}, which was later named Copilot.

Figure \ref{img:new_bing} provides a visual representation of New Bing's search interface architecture. The left panel displays conventional search results, each comprising three standard components: the website title, corresponding hyperlink, and an algorithmically selected excerpt deemed most relevant to the search query. Meanwhile, the right panel features Bing Chat, which generates a response by integrating its knowledge base with real-time information extracted from relevant web sources, functionally analogous to Google's AI Overview. Source citations appear as hyperlinks positioned at the bottom of the response interface, with some sources receiving explicit in-text citations, while others are listed as supplementary references without direct textual attribution. Consistent with Google AI Overview's citation patterns, a single sentence may incorporate zero to multiple source references, while each website may support zero to multiple sentences within the generated response.

\begin{figure}[H]
\FIGURE{\includegraphics[width=6in]{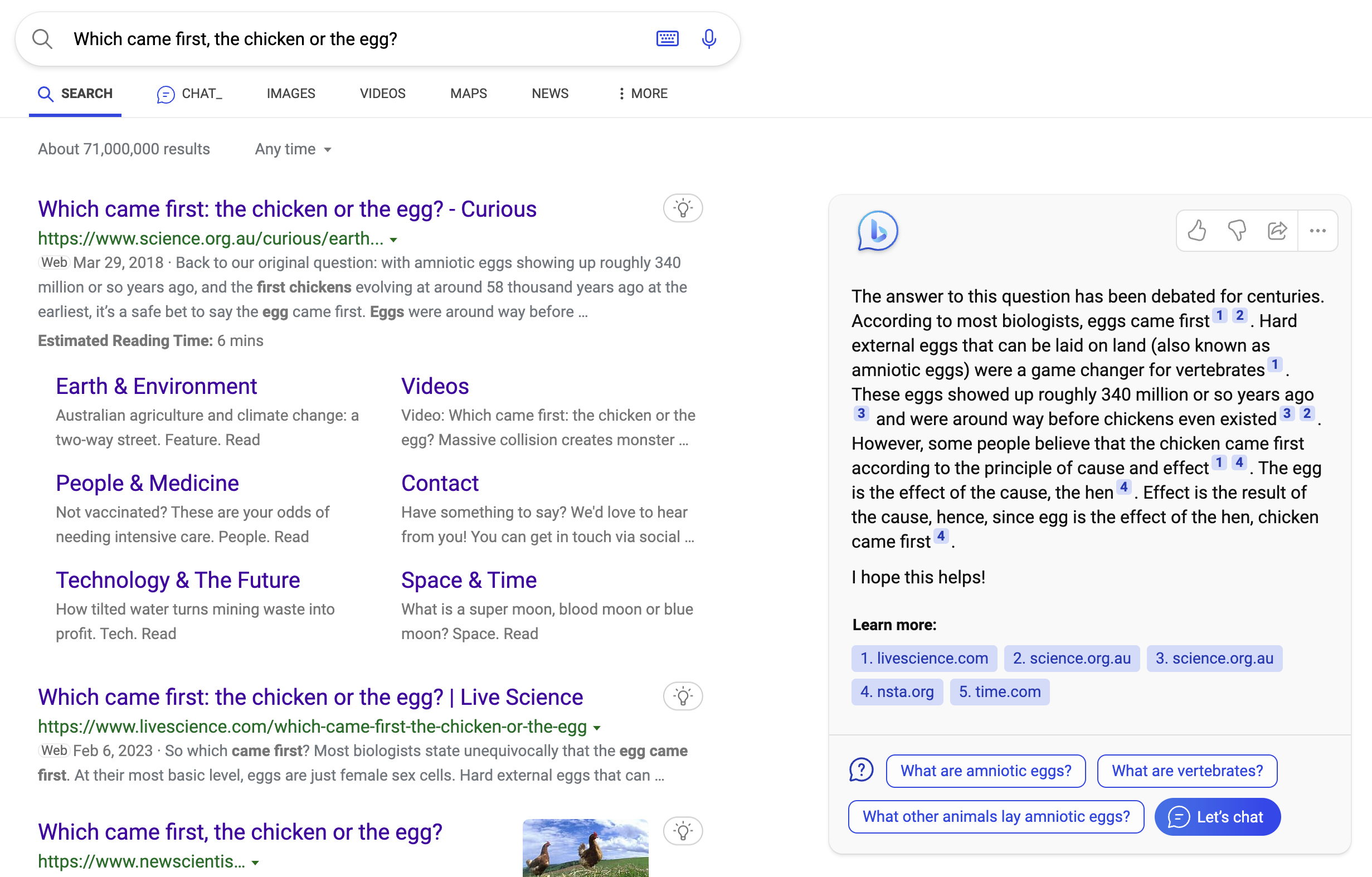}}
{An Example of Using New Bing \label{img:new_bing}}
{}
\end{figure}

In April 2023, following the same procedure established for our primary Google analyses, we constructed a dataset by randomly sampling queries from HC3 \citep{guo2023close}. Given the absence of publicly available APIs for New Bing, we stop the collection process after successfully executing 700 queries, capturing both the top 20 organic search results from the conventional search panel and the corresponding AI-generated content from the chat interface. We also downloaded all referenced website content (13,428 in total) from both channels for analysis.

To evaluate the robustness of our RQ1 findings, we implement an identical analytical procedure for the New Bing dataset. Specifically, we apply equivalent chunk selection processes and calculate both perplexity and semantic similarity measures following our established methodology. The sole methodological modification involves substituting GPT-2 for Google's Gemma-2B model in perplexity calculations, given that the GPT series serves as the foundational architecture underlying Microsoft's Bing Chat system. Our replication analysis confirms the consistency of our findings to RQ1: websites cited by Bing Chat exhibit significantly lower perplexity compared to those exclusively listed in traditional Bing search results, while perplexity measures demonstrate no significant predictive power for traditional Bing rankings. Additionally, we observe that Bing Chat's content selections demonstrate significantly greater semantic similarity compared to sources prioritized by conventional Bing search algorithms.

\begin{table}[!h]
\centering
\caption{Perplexity Effect (Bing)}
\label{tab:Bing_perplex}
\newcolumntype{L}[1]{>{\raggedright\arraybackslash}p{#1}}
\newcolumntype{C}[1]{>{\centering\arraybackslash}p{#1}}
\newcolumntype{R}[1]{>{\raggedleft\arraybackslash}p{#1}}
\renewcommand\arraystretch{0.5}
\begin{tabular}{L{2.5cm} c c c c}
\hline\hline
& \multicolumn{2}{c}{\(ChatCite\)} & \multicolumn{2}{c}{\(SentenceCite\)} \\
\cline{2-5}
& \(LPM\) & \(Logit\) & \(LPM\) & \(Logit\) \\
\cline{2-5}
& (1) & (2) & (3) & (4) \\
\hline
\(PPL\) 
  & -0.0755*** & -0.4722*** & -0.0443*** & -0.6343*** \\
& (0.0119) & (0.0670) & (0.0050) & (0.0521) \\
\hline
Query FE    & Yes & Yes & Yes & Yes \\
Observations     & 13,428 & 13,428 & 49,917  & 49,917 \\
\(R^2\) or \(Chi^2\)    & 0.03   & 49.24   & 0.05   & 143.71 \\
\hline\hline
\end{tabular}
\begin{tablenotes}
\footnotesize
\centerline{Note: Cluster-robust standard errors are reported at the query level. *** $p$ $<$ 0.01, ** $p$ $<$ 0.05, * $p$ $<$ 0.1.}
\end{tablenotes}
\end{table}

\begin{table}[!h]
\centering
\caption{Semantic Similarity Effect (Bing)}
\label{tab:bing_simi}
\newcolumntype{L}[1]{>{\raggedright\arraybackslash}p{#1}}
\newcolumntype{C}[1]{>{\centering\arraybackslash}p{#1}}
\newcolumntype{R}[1]{>{\raggedleft\arraybackslash}p{#1}}
\renewcommand\arraystretch{0.5}
\begin{tabular}{L{3cm} c c}
\hline\hline
& \multicolumn{2}{c}{\(Similarity\)} \\
\cline{2-3}
 & (1) & (2) \\
\hline
\(BothCite\) 
  & 0.0934*** & 0.0901*** \\
& (0.0052) & (0.0015) \\
\hline
Cross-Category     & No & Yes\\
Query FE    & Yes & Yes \\
Observations     & 83,923 & 122,568\\
\(R^2\)          & 0.42   & 0.41 \\
\hline\hline
\end{tabular}
\begin{tablenotes}
\footnotesize
\centering
Note: Cluster-robust standard errors are reported at the query level. *** $p$ $<$ 0.01, ** $p$ $<$ 0.05, * $p$ $<$ 0.1; \\ ``Cross-Category'' indicates whether $BothCite = 0$ includes pairs with one cited chunk and one non-cited chunk.
\end{tablenotes}
\end{table}

To validate the robustness of our finding that observed citation preferences emerge intrinsically from underlying language model architectures rather than from platform-specific engineering modifications, we conduct an additional series of LLM-based RAG experiments. In this validation study conducted in July 2025, we substitute Google's Gemini API with OpenAI's GPT-4o-mini API, maintaining architectural consistency with New Bing's GPT-series foundation during our observation period\footnote{https://blogs.microsoft.com/blog/2023/02/07/reinventing-search-with-a-new-ai-powered-microsoft-bing-and-edge-your-copilot-for-the-web/}. Our empirical analyses corroborate the generalizability of our findings: content selected by the GPT-based RAG system exhibits significantly lower perplexity and demonstrates substantially greater semantic homogeneity compared to non-selected content. Additionally, we note that we employ the deprecated GPT-4 for analyses conducted in January 2024 and obtain similar results, further substantiating the generalizability of our findings.

\begin{table}[!h]
\centering
\caption{RAG's Perplexity Effect and Position Bias (GPT)}
\label{tab:rag_perplex_gpt}
\newcolumntype{L}[1]{>{\raggedright\arraybackslash}p{#1}}
\newcolumntype{C}[1]{>{\centering\arraybackslash}p{#1}}
\newcolumntype{R}[1]{>{\raggedleft\arraybackslash}p{#1}}
\renewcommand\arraystretch{0.5}
\begin{tabular}{L{2.5cm} c c c c}
\hline\hline
& \multicolumn{2}{c}{\(RAGCite\)}\\
\cline{2-3}
& \(LPM\) & \(Logit\) \\
\cline{2-3}
& (1) & (2)\\
\hline
\(PPL\) 
  & -0.0321*** & -0.156*** \\
& (0.0084) & (0.0310)  \\
\(Pos\) 
  & -0.0163*** & -0.0644*** \\
& (0.0008) & (0.0034)  \\
\hline
Query FE    & Yes & Yes \\
Observations     & 13,428 & 13,428  \\
\(R^2\) or \(Chi^2\)   & 0.11   & 500.55\\
\hline\hline
\end{tabular}
\begin{tablenotes}
\footnotesize
\centerline{Note: Cluster-robust standard errors are reported at the query level. *** $p$ $<$ 0.01, ** $p$ $<$ 0.05, * $p$ $<$ 0.1.}
\end{tablenotes}
\end{table}

\begin{table}[!h]
\centering
\caption{RAG's Semantic Similarity Effect (GPT)}
\label{tab:rag_simi_gpt}
\newcolumntype{L}[1]{>{\raggedright\arraybackslash}p{#1}}
\newcolumntype{C}[1]{>{\centering\arraybackslash}p{#1}}
\newcolumntype{R}[1]{>{\raggedleft\arraybackslash}p{#1}}
\renewcommand\arraystretch{0.5}
\begin{tabular}{L{3cm} c c}
\hline\hline
& \multicolumn{2}{c}{\(Similarity\)} \\
\cline{2-3}
 & (1) & (2) \\
\hline
\(BothRAGCite\) 
  & 0.0992*** & 0.0769*** \\
& (0.0035) & (0.0028) \\
\hline
Cross-Category     & No & Yes\\
Query FE    & Yes & Yes \\
Observations     & 67,410 & 122,568\\
\(R^2\)          & 0.45   & 0.42 \\
\hline\hline
\end{tabular}
\begin{tablenotes}
\footnotesize
\centerline{Note: Cluster-robust standard errors are reported at the query level. *** $p$ $<$ 0.01, ** $p$ $<$ 0.05, * $p$ $<$ 0.1.}
\end{tablenotes}
\end{table}

Finally, we conduct a robustness validation for RQ3, examining how LLM-based content polishing influences generative search outputs. We implement the identical experimental protocol as our primary Google analyses, substituting GPT-4o-mini for Gemini in the content refinement process. Our polishing prompts also specify optimization for Bing Chat appearance to maintain platform-specific relevance. The robustness analysis corroborates our principal findings: content optimization results in expanded source citation by the RAG system, further enhancing the information diversity of generated outputs.

\begin{table}[H]
\centering
\caption{Impact of Sources' LLM Polishing on (Cited) Content's Similarity (GPT)}
\label{tab:polish_simi_gpt}
\newcolumntype{L}[1]{>{\raggedright\arraybackslash}p{#1}}
\newcolumntype{C}[1]{>{\centering\arraybackslash}p{#1}}
\newcolumntype{R}[1]{>{\raggedleft\arraybackslash}p{#1}}
\renewcommand\arraystretch{0.5}
\begin{tabular}{L{3cm} c c}
\hline\hline
& \multicolumn{2}{c}{\(Similarity\)} \\
\cline{2-3}
 & (1) & (2) \\
\hline
\(T = 1\) 
  & -0.0083*** & -0.0284*** \\
& (0.0006) & (0.0009) \\
\(T = 2\) 
  & -0.0078*** & -0.0657*** \\
& (0.0004) & (0.0013) \\
\hline
Sample     & All & Cited Only\\
Query FE    & Yes & Yes \\
Observations     & 367,704 & 50,933\\
\(R^2\)          & 0.24   & 0.40 \\
\hline\hline
\end{tabular}
\begin{tablenotes}
\footnotesize
\centerline{Note: Cluster-robust standard errors are reported at the query level. *** $p$ $<$ 0.01, ** $p$ $<$ 0.05, * $p$ $<$ 0.1.}
\end{tablenotes}
\end{table}

\begin{table}[H]
\centering
\caption{Impact of Sources' LLM Polishing on Outputs' Attributes (GPT)}
\label{tab:polish_output_gpt}
\newcolumntype{L}[1]{>{\raggedright\arraybackslash}p{#1}}
\newcolumntype{C}[1]{>{\centering\arraybackslash}p{#1}}
\newcolumntype{R}[1]{>{\raggedleft\arraybackslash}p{#1}}
\renewcommand\arraystretch{0.5}
\begin{tabular}{L{3cm} c c}
\hline\hline
& \(NumCite\) & \(OutputPPL\) \\
\cline{2-3}
 & (1) & (2) \\
\hline
\(T = 1\) 
  & 0.7820*** & -0.0024 \\
& (0.0916) & (0.0030) \\
\(T = 2\) 
  & 1.8144*** & -0.0035 \\
& (0.2179) & (0.0041) \\
\hline
Observations     & 2,100 & 2,100\\
\(R^2\)          & 0.04 & 0.01 \\
\hline\hline
\end{tabular}
\begin{tablenotes}
\footnotesize
\centerline{Note: Cluster-robust standard errors are reported at the query level. *** $p$ $<$ 0.01, ** $p$ $<$ 0.05, * $p$ $<$ 0.1.}
\end{tablenotes}
\end{table}

\newpage
\section{Outlier Removal}
\label{appendix: outlier}

Some websites may contain atypical content that exhibits excessive unpredictability from the LLM's perspective, and these outliers may constitute the primary driver of the observed perplexity effects. Therefore, we exclude content chunks with the top 1\% perplexity values and replicate our analyses using the same specifications as Equations \ref{eq:1} and \ref{eq:2}. As demonstrated in Table \ref{tab:outlier}, our results remain robust to the removal of such outlier websites.

\begin{table}[H]
\centering
\caption{Perplexity Effect after Removing Outliers}
\label{tab:outlier}
\newcolumntype{L}[1]{>{\raggedright\arraybackslash}p{#1}}
\newcolumntype{C}[1]{>{\centering\arraybackslash}p{#1}}
\newcolumntype{R}[1]{>{\raggedleft\arraybackslash}p{#1}}
\renewcommand\arraystretch{0.5}
\begin{tabular}{L{2.5cm} c c c c}
\hline\hline
& \multicolumn{2}{c}{\(ChatCite\)} & \multicolumn{2}{c}{\(SentenceCite\)} \\
\cline{2-5}
& \(LPM\) & \(Logit\) & \(LPM\) & \(Logit\) \\
\cline{2-5}
& (1) & (2) & (3) & (4) \\
\hline
\(PPL\) 
  & -0.0118*** & -0.0550*** & -0.0021*** & -0.0621*** \\
& (0.0002) & (0.0010) & (0.0001) & (0.0016) \\
\hline
Query FE    & Yes & Yes & Yes & Yes \\
Observations     & 97,528 & 97,528 & 250,844 & 250,844 \\
\(R^2\) or \(Chi^2\)     & 0.14 & 3519.26 & 0.06  &  1807.68 \\
\hline\hline
\end{tabular}
\begin{tablenotes}
\footnotesize
\centerline{Note: Cluster-robust standard errors are reported at the query level. *** $p$ $<$ 0.01, ** $p$ $<$ 0.05, * $p$ $<$ 0.1.}
\end{tablenotes}
\end{table}

\newpage
\section{Alternative Samples}
\label{appendix: samples}

In our dataset, the average number of websites retrieved from conventional search engines exceeds that from generative search engines. This disparity may partially account for the relatively higher information diversity observed in conventional search results, as the larger sample size could influence the findings. To control for this potential confound, we implement a balanced sampling approach: for each query $q$, we utilize the minimum website count across both channels as an upper limit, ensuring equivalent sample sizes for comparative analysis. Using this balanced sample, we compare the semantic similarity between cited and non-cited website pairs employing the same specification as Equation \ref{eq:3}. As demonstrated in Table \ref{tab:simi_balanced}, our results remain qualitatively unaffected by this sample balancing process. This robustness confirms that the observed differences in information diversity are not merely artifacts of unequal sample sizes between search channels.

\begin{table}[H]
\centering
\caption{Semantic Similarity Effect with Balanced Sample}
\label{tab:simi_balanced}
\newcolumntype{L}[1]{>{\raggedright\arraybackslash}p{#1}}
\newcolumntype{C}[1]{>{\centering\arraybackslash}p{#1}}
\newcolumntype{R}[1]{>{\raggedleft\arraybackslash}p{#1}}
\renewcommand\arraystretch{0.5}
\begin{tabular}{L{3cm} c}
\hline\hline
& {\(Similarity\)} \\
\hline
\(BothCite\) 
  & 0.0355*** \\
& (0.0020) \\
\hline
Query FE    & Yes \\
Observations     & 350,944\\
\(R^2\)          & 0.28 \\
\hline\hline
\end{tabular}
\begin{tablenotes}
\footnotesize
\centering
Note: Cluster-robust standard errors are reported at the query level. *** $p$ $<$ 0.01, ** $p$ $<$ 0.05, * $p$ $<$ 0.1.
\end{tablenotes}
\end{table}

\newpage
\section{Alternative Specifications}
\label{appendix: specification}

In our semantic similarity analyses, the unit of observation comprises individual pairs of website chunks. Here, we consider an alternative data organization approach and specify corresponding econometric analyses. Specifically, for each query $q$, we calculate the average semantic similarity across all pairs of cited chunks and denote this value as $AvgSimilarity_{q, 1}$. Simultaneously, we calculate the average semantic similarity across all pairs of non-cited chunks and denote this value as $AvgSimilarity_{q, 0}$. We conduct a direct pairwise t-test between these two sets of values and obtain consistent results: cited chunks from the generative engine exhibit significantly greater mutual similarity compared to those not cited by the generative engine.

\newpage
\section{Perplexity Effect in Conventional Search Engines}
\label{appendix:ppl_conventional}

Conventional search rankings may also select websites with lower perplexity, potentially establishing a baseline that could confound our results regarding generative engines. Although the overlap between the two channels' selections is minimal in our dataset, we directly test this potential confounding factor. Specifically, using websites ranked by conventional search engines, we examine whether the perplexity of a website chunk $PPL_{q,c}$ predicts the ranking of the website $Rank_{q,c}$ using the following specification:
\begin{equation}
\label{eq:ppl_conv}
    Rank_{q,c} = \beta_0 + \beta_1 * PPL_{q,c} + Q_q + \epsilon_{q,c}
\end{equation}

As demonstrated below, the perplexity effect is absent in conventional ranking mechanisms, thereby eliminating this potential confounding factor from our main results.
\begin{table}[H]
\centering
\caption{Perplexity Effect (Conventional Search)}
\label{tab:ppl_conv}
\newcolumntype{L}[1]{>{\raggedright\arraybackslash}p{#1}}
\newcolumntype{C}[1]{>{\centering\arraybackslash}p{#1}}
\newcolumntype{R}[1]{>{\raggedleft\arraybackslash}p{#1}}
\renewcommand\arraystretch{0.5}
\begin{tabular}{L{3cm} c}
\hline\hline
& {\(Rank\)} \\
\hline
\(PPL\) 
  & 0.0144 \\
& (0.0201) \\
\hline
Query FE    & Yes \\
Observations     & 57,653\\
\(R^2\)          & 0.07 \\
\hline\hline
\end{tabular}
\begin{tablenotes}
\footnotesize
\centering
Note: Cluster-robust standard errors are reported at the query level. *** $p$ $<$ 0.01, ** $p$ $<$ 0.05, * $p$ $<$ 0.1.
\end{tablenotes}
\end{table}

\newpage
\section{Experimental Design}
\label{appendix: experiment}

\subsection{Overview and Participant Flow}
\begin{figure}[h]
    \caption{Experimental workflow diagram. Participants were randomly assigned to either original website content (control) or LLM-polished content (treatment) conditions through deterministic assignment based on Prolific ID.}
    \centering
    \includegraphics[width=0.75\linewidth]{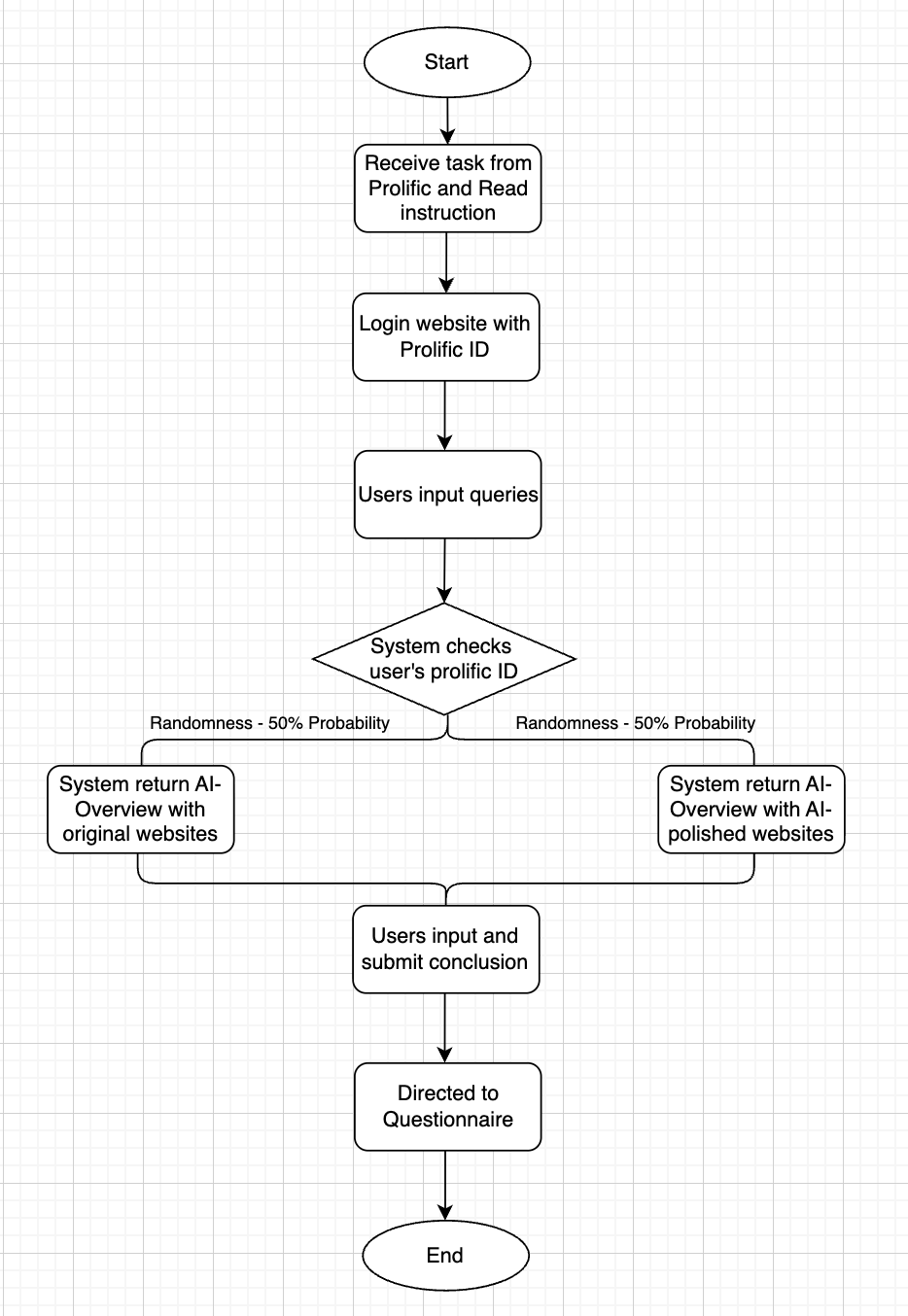}

    \label{fig:flowchart}
\end{figure}

We develop a custom web-based generative search platform that replicates the core functionality of Google AI Overview while keeping track of users' activities.  Figure~\ref{fig:flowchart} illustrates the complete experimental workflow. Participants proceed through six sequential stages: initial task receipt, platform authentication, iterative search interactions, policy brief composition, submission, and post-task questionnaire completion.

\subsection{Initial Landing Page and Authentication}

Participants are recruited from Prolific. Upon accepting the study, participants receive a direct link to our experimental platform. The initial landing page (Figure~\ref{fig:welcome-login}) presents participants with task instructions alongside the authentication interface. Participants are instructed to assume the role of education policy consultants tasked with developing a policy recommendation regarding smartphone bans in K-12 educational settings. The task requires participants to research and synthesize information to produce a policy brief with 100-300 words. 

After reading the task instructions, if participants decide to take the task, we require them to enter their unique Prolific ID for authorization. Getting the unique Prolific ID of each participant serves us dual purposes: identity verification and deterministic treatment assignment. 

\begin{figure}[h]
    \caption{Platform authentication and instruction interface. Participants review task requirements before entering their Prolific ID to access the search system.}
    \centering
    \includegraphics[width=0.75\linewidth]{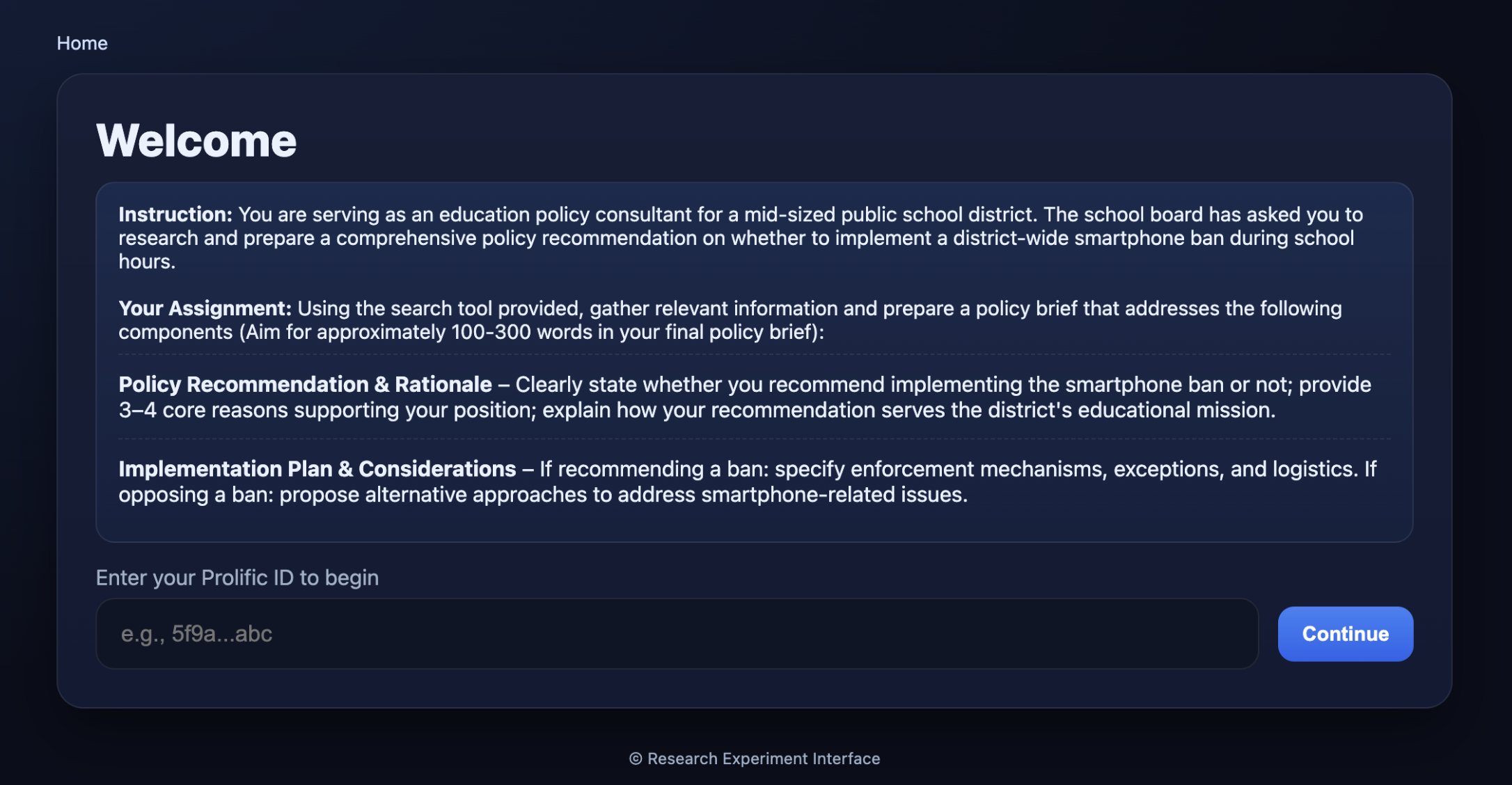}
    \label{fig:welcome-login}
\end{figure}

\subsection{Treatment Assignment Mechanism}
Following successful authentication, our system randomly assigns each participant to either the treatment group or the control group through a hash-based assignment algorithm. To ensure stability across page reloads and repeated sessions, the server computes the \texttt{Prolific ID} and maps it to {control, treatment}. This approach ensures balanced allocation (approximately 50\% per condition) while maintaining assignment stability across sessions, as we allow our participants to search multiple times as they would like. The assignment into treatment or control groups determines which website content pool would underlie all their subsequent AI-generated responses. Control group participants interact with a RAG system that accesses 65 original, unmodified website passages extracted from Google and Bing search results related to smartphone policies in educational settings. Treatment group participants, in contrast, receive AI-generated responses based on the same 65 website passages after LLM-based polishing with citation optimization objectives, as detailed in Section~\ref{sec: LLM_polish} of the manuscript. This manipulation of the source websites represents the sole experimental variation between our treatment and control groups. All interface elements, interaction mechanisms, and system functionalities remain identical between treatment groups, isolating the effect of content polishing on user behavior and outcomes.

\subsection{Search Interface and AI Overview Generation}

The primary experimental interface (Figure~\ref{fig:AI_Overview_Panel}) consists of three integrated components: a search query field, an AI-generated overview panel, and a response composition area. Participants can submit natural language queries through the search field, triggering a server-side processing sequence that begins with query receipt and verification of the participant's treatment or control condition assignment. The system then retrieves relevant information from the assigned source website pool, either original or polished content, depending on the experimental condition, and generates an AI summary using Gemini's RAG API. In addition to the summary generation, the system also extracts and formats citations from the synthesized content, ultimately delivering the response with inline numeric citations to the participant's interface.

The generated AI Overview appears immediately below the search field, presenting synthesized information with embedded citations formatted as bracketed numbers (e.g. [1][2][5]). Each citation links to its corresponding source website content, which participants could access in new browser tabs to examine original content while maintaining their workflow in the main interface. The system utilizes AJAX requests to refresh only the AI Overview panel during searches, preserving any text that participants have already composed in the submission area.

\begin{figure}[!h]
\caption{AI Overview interface displaying generated summary with inline citations and linked source documents. }
    \centering
    \includegraphics[width=0.75\linewidth]{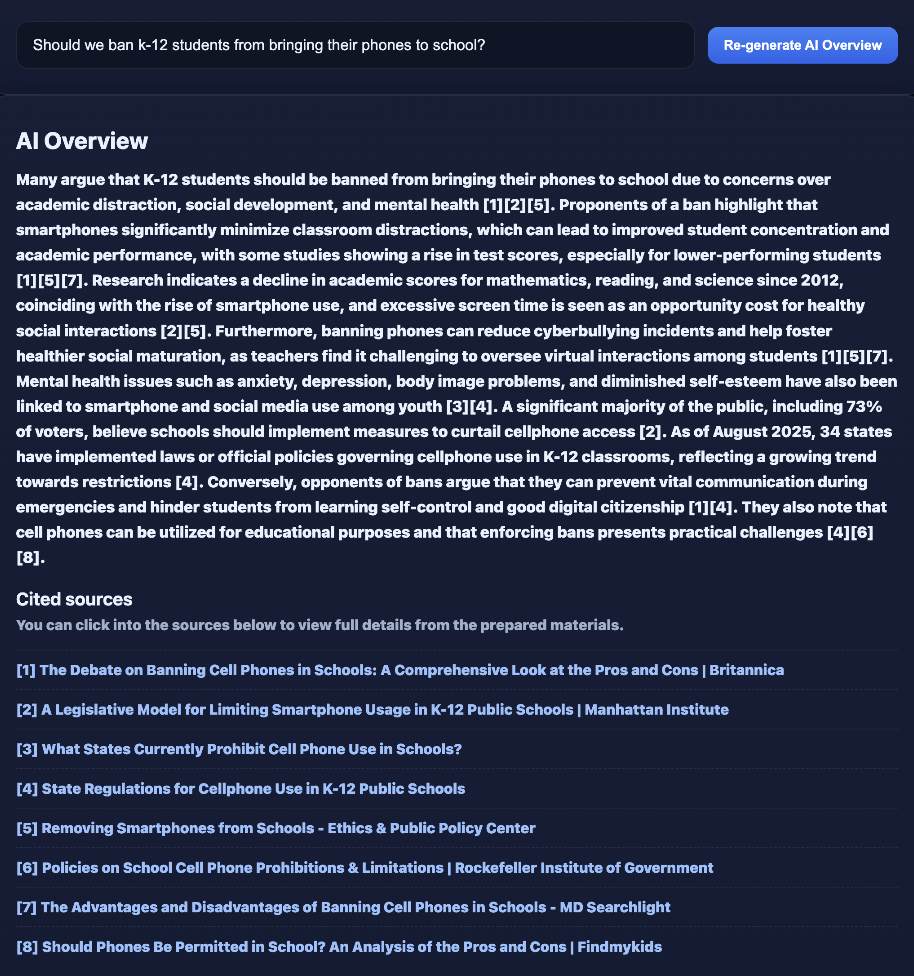}
    \label{fig:AI_Overview_Panel}
\end{figure}

\subsection{Response Composition and Submission}

Below the AI Overview panel, participants compose their policy briefs in a text area (Figure~\ref{fig:submit}). The interface enforces two constraints to ensure task engagement and quality. First, a minimum length requirement ensures that the submission button is disabled until participants compose at least 100 words, with the interface displaying real-time word count feedback. Second, we prohibit paste operations into both query and response fields, requiring participants to actively synthesize information rather than mechanically reproduce AI-generated content. Participants can conduct unlimited searches throughout the task, with the system logging all queries, generated responses, citation patterns, and interaction timestamps. When participants initiate new searches, the system preserves any text they have already composed in the response field, enabling them to iteratively refine their policy briefs while exploring different aspects of the smartphone ban debate without losing their work in progress.

\begin{figure}[!h]
    \caption{Policy brief Composition and Submission: the system enforced a 100-word minimum while preventing mechanical content reproduction through paste restrictions.}
    \centering
    \includegraphics[width=0.75\linewidth]{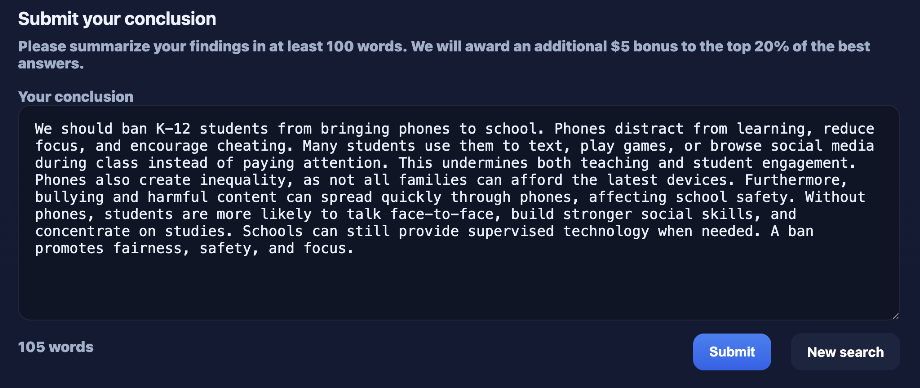}

    \label{fig:submit}
\end{figure}

\subsection{Data Collection and Post-Task Assessment}
Upon response submission, our system captures and stores a comprehensive set of session data, including final responses, completion timestamps, search histories, and interaction patterns. Participants then proceed automatically to a post-task questionnaire (detailed in Appendix~\ref{appendix: survey})  that collects information across two primary dimensions. First, it evaluates their experimental task experience using the well-established constructs from technology acceptance research, the perceived ease of use and the perceived usefulness, each measured through three validated questions. Second, the questionnaire assesses participants' prior experience with relevant technologies through frequency-of-use measures: their ChatGPT usage and search engine usage (Google, Bing, or other platforms) over the preceding four weeks. Of the 150 participants initially recruited, three failed to complete the post-task questionnaire and were excluded from analyses, yielding a final sample of 147 participants with complete experimental data.

\newpage
\section{Post-Task Survey Questionnaire}
\label{appendix: survey}

Please indicate your experience and perceptions when you used the provided tool for completing the assigned tasks. 
The first six items were measured on a 7-point Likert scale (1 = strongly disagree, 7 = strongly agree).

\surveysection{Perceived ease of use}
\begin{enumerate}
    \item I find completing the tasks using the provided tool easy.
    \item I feel skillful in doing the tasks by the provided tool.
    \item My interaction with the provided tool is clear and understandable.
  
\end{enumerate}

\surveysection{Perceived usefulness}
\begin{enumerate}
    \item Using the provided tool enabled me to accomplish the tasks quickly.
    \item Using the provided tool enhanced my effectiveness in completing the tasks.
    \item I find the provided tool useful.
\end{enumerate}

\surveysection{Tool Experience}
\begin{enumerate}
    \item In the past 4 weeks, about how often did you use ChatGPT: Never in the past 4 weeks; Less than once per week (~1–3 times in the past 4 weeks); About once per week; 2–3 times per week; 4–6 times per week; About once per day (~7 × per week); Several times per day.
    \item In the past 4 weeks, about how often did you use search engines (Google/Bing/etc): Never in the past 4 weeks; Less than once per week (~1–3 times in the past 4 weeks); About once per week; 2–3 times per week; 4–6 times per week; About once per day (~7 × per week); Several times per day.
\end{enumerate}

\surveysection{Education}
\begin{enumerate}
    \item What is your educational background: High School or Below; Some College Credit, No Degree; Bachelor's Degree; Master's Degree; Doctorate Degree
    
\end{enumerate}

\newpage
\section{Randomization Checks and Other Comparisons}
\label{appendix: randomization}

Based on data collected from Prolific participants, we conduct pairwise t-tests across experimental groups on gender (binary coded), age, and education (binary coded for college degree or higher). Given that the majority of participants identify as White, we construct a binary ethnicity variable (White versus non-White) and conduct a pairwise t-test on this variable as well. Additionally, since we collect data on participants' prior experience with generative AI and search engines, which constitute categorical variables, we employ Kolmogorov-Smirnov tests for these measures. These analyses collectively confirm that participants across treatment and control groups are balanced, indicating successful randomization.

Beyond randomization checks, we also examine differences in participants' perceived ease of use and perceived usefulness measures collected through our survey. However, we find no statistically significant differences between groups on these dimensions.

\begin{table}[H]
\centering
\caption{Tests Results}
\label{tab:balance_tests}
\newcolumntype{L}[1]{>{\raggedright\arraybackslash}p{#1}}
\newcolumntype{C}[1]{>{\centering\arraybackslash}p{#1}}
\newcolumntype{R}[1]{>{\raggedleft\arraybackslash}p{#1}}
\renewcommand\arraystretch{0.5}
\begin{tabular}{L{4cm} c c}
\hline\hline
Variables & \(t\)/Combined \(K-S\) Statistics & \(p\)-value \\
\hline
\(Gender\) & 1.42  & 0.16 \\
\(Age\) & -1.60 & 0.11 \\
\(College\) & 0.25  & 0.80 \\
\(White\) & -0.27 & 0.79 \\
\(Generative\_AI\_Usage\) & 0.05  & 1.00 \\
\(Search\_Engine\_Usage\) & 0.09 & 0.93 \\
\(Ease\_to\_Use\) & 1.12  & 0.27 \\
\(Usefulness\) & -1.18 & 0.24 \\
\hline\hline
\end{tabular}
\begin{tablenotes}
\footnotesize
\centering
Note: *** $p$ $<$ 0.01, ** $p$ $<$ 0.05, * $p$ $<$ 0.1.
\end{tablenotes}
\end{table}

\newpage
\section{Vendi Score Calculation}
\label{appendix: vendi}

Let a submission (document) $d$ be segmented into $n$ text units (in our case, sentences)
$\mathcal{S}_d=\{s_1,\ldots,s_n\}$. The \emph{Vendi score} provides an
``effective-number'' measure of how many mutually distinct items are present in $\mathcal{S}_d$,
based on the spectrum of a similarity kernel.

\paragraph{Step 1: Segment and embed.}
Tokenize $d$ into sentences and produce vector representations
$\mathbf{e}_i \in \mathbb{R}^p$ for each $s_i$ using a sentence embedding model (SBERT).
L2-normalize embeddings so $\|\mathbf{e}_i\|_2=1$.

\paragraph{Step 2: Build a PSD similarity kernel.}
Form the $n\times n$ Gram matrix $K$ with entries
\begin{equation}
K_{ij} \;=\; k(\mathbf{e}_i,\mathbf{e}_j),
\qquad
\text{e.g., } k(\mathbf{e}_i,\mathbf{e}_j)=\mathbf{e}_i^\top \mathbf{e}_j
\text{ (cosine)} \text{ or } k(\mathbf{e}_i,\mathbf{e}_j)=\exp(-\gamma\|\mathbf{e}_i-\mathbf{e}_j\|_2^2)
\text{ (RBF)}.
\label{eq:kernel}
\end{equation}
To ensure unit diagonal (required for the interpretation below), rescale
\begin{equation}
\tilde K \;=\; D^{-1/2} K D^{-1/2},
\qquad
D \;=\; \mathrm{diag}\big(K_{11},\ldots,K_{nn}\big).
\label{eq:unitdiag}
\end{equation}

\paragraph{Step 3: Normalize and take the spectrum.}
Let $\{\lambda_i\}_{i=1}^n$ be the eigenvalues of $\tilde K / n$.
Because $\tilde K$ is positive semidefinite and diagonally normalized, the
$\lambda_i$ are nonnegative and (up to numerical tolerance) sum to one.

\paragraph{Step 4: Entropy of the spectrum and the Vendi score.}
Define the spectral (Shannon) entropy
\begin{equation}
H(\tilde K) \;=\; - \sum_{i=1}^n \lambda_i \log \lambda_i,
\label{eq:spectral-entropy}
\end{equation}
where $\log$ is the natural logarithm.
The \emph{Vendi score} for submission $d$ is the corresponding Hill number
\begin{equation}
\mathrm{VS}(d) \;=\; \exp\!\big( H(\tilde K) \big)
\;\in\; [1,\,n].
\label{eq:vendi}
\end{equation}

\paragraph{Interpretation and edge cases.}
If all items are identical (perfect similarity), then $\tilde K$ is rank one,
$H(\tilde K)=0$, and $\mathrm{VS}(d)=1$. If items are mutually orthogonal
(e.g., $\tilde K=I_n$), then $\lambda_i=\tfrac{1}{n}$, $H(\tilde K)=\log n$,
and $\mathrm{VS}(d)=n$. Thus, $\mathrm{VS}(d)$ can be read as the
\emph{diversity-equivalent} number of distinct segments in $d$.

\end{APPENDICES}

\end{document}